\RequirePackage[2014/01/01]{latexrelease}

\documentclass[acmsmall,nonacm]{acmart}

\setcopyright{none}

\usepackage[T1]{fontenc}
\usepackage{amsmath}
\usepackage{fancybox}
\usepackage{multirow}
\usepackage{subcaption}
\usepackage{mathrsfs}
\usepackage{mathtools}
\usepackage{xspace}

\newcommand{\pagelength}{w}
\usepackage{amsthm}
\usepackage{listings}
\usepackage{amssymb}
\usepackage{amsfonts}
\usepackage{gdefs}
\usepackage{setspace}
\usepackage{hhline}
\usepackage{refs}
\usepackage{relsize}
\usepackage{stmaryrd}
\usepackage{algorithm}
\usepackage[noend]{algpseudocode}
\usepackage{mathdots}
\usepackage{kbordermatrix}
\usepackage{stackrel}
\usepackage{alltt}
\usepackage{sidecap}
\makeatletter 
\AtEndDocument{%
\def\cb@checkPdfxy#1#2#3#4#5{%
\cb@@findpdfpoint{#1}{#2}%
\ifdim#3sp=\cb@pdfx\relax      
\ifdim#4sp=\cb@pdfy\relax      
\ifdim#5=\cb@pdfz\relax
\else
\cb@error
\fi
\else
\cb@error
\fi
\else
\cb@error
\fi
}}
\makeatother 

\usepackage{color}
\usepackage{mathpartir} 
\usepackage{tikz}
\usetikzlibrary{shapes,arrows}

\usepackage[color]{changebar}
\cbcolor{red}

\newcommand{\OnlySubmission}[1]{}
\newcommand{\OnlyConference}[1]{}
\newcommand{\OnlyTech}[1]{}

{\par\noindent{\bf Theorem~\ref{The:#1}.\/}\begin{it}}%
	{\end{it}}
{\par\noindent{\bf Lemma~\ref{Lem:#1}.\/}\begin{it}}%
	{\end{it}}
{\par\noindent{\bf Corollary~\ref{Cor:#1}.\/}\begin{it}}%
	{\end{it}}

\usepackage{enumitem}
\setlist{nosep,leftmargin=\parindent}

\AtBeginDocument{%
\setlength\abovedisplayskip{0.5\abovedisplayskip}%
\setlength\belowdisplayskip{0.5\belowdisplayskip}%
\setlength\abovedisplayshortskip{0.5\abovedisplayshortskip}%
\setlength\belowdisplayshortskip{0.5\belowdisplayshortskip}%
\setlength\floatsep{0.5\floatsep}%
}

\newcounter{LineNumber}[figure]

\newcommand{\toolname}{\textrm{GenXGen[MC]}\xspace}
\lccode`\[`\[
\lccode`\]`\]
\newcommand{\toolnameC}{\textrm{GenXGen[C]}\xspace}

\newcommand{\PE}{\textit{pe}}
\newcommand{\GE}{\textit{ge}}
\newcommand{\assignment}[1]{A(#1)}

\newcommand{\Omit}[1]{}

\makeatletter
\newsavebox{\@brx}
\newcommand{\llangle}[1][]{\savebox{\@brx}{\(\m@th{#1\langle}\)}%
	\mathopen{\copy\@brx\kern-0.5\wd\@brx\usebox{\@brx}}}
\newcommand{\rrangle}[1][]{\savebox{\@brx}{\(\m@th{#1\rangle}\)}%
	\mathclose{\copy\@brx\kern-0.5\wd\@brx\usebox{\@brx}}}
\makeatother

\newcommand{\subsubsubsection}[1]{\vspace{2pt plus 1pt minus 1pt}\noindent{\bf #1}}

\newcommand{\asm}[1]{\texttt{#1}}

\begin{document}

\title[Generating Extension Generator]{A Generating-Extension-Generator for
  Machine Code
}


\author{Michael Vaughn}
\affiliation{
  \institution{University of Wisconsin}
}
\email{mvaughn@wisc.edu}

\author{Thomas Reps}
\affiliation{%
	\institution{University of Wisconsin}
}
\email{reps@cs.wisc.edu}

\Omit{
\author{First1 Last1}
\authornote{with author1 note}          
\orcid{nnnn-nnnn-nnnn-nnnn}             
\affiliation{
  \position{Position1}
  \department{Department1}              
  \institution{Institution1}            
  \streetaddress{Street1 Address1}
  \city{City1}
  \state{State1}
  \postcode{Post-Code1}
  \country{Country1}                    
}
\email{first1.last1@inst1.edu}          

\author{First2 Last2}
\authornote{with author2 note}          
\orcid{nnnn-nnnn-nnnn-nnnn}             
\affiliation{
  \position{Position2a}
  \department{Department2a}             
  \institution{Institution2a}           
  \streetaddress{Street2a Address2a}
  \city{City2a}
  \state{State2a}
  \postcode{Post-Code2a}
  \country{Country2a}                   
}
\email{first2.last2@inst2a.com}         
\affiliation{
  \position{Position2b}
  \department{Department2b}             
  \institution{Institution2b}           
  \streetaddress{Street3b Address2b}
  \city{City2b}
  \state{State2b}
  \postcode{Post-Code2b}
  \country{Country2b}                   
}
\email{first2.last2@inst2b.org}         
}


\begin{abstract}
  The problem of ``debloating'' programs for security and performance purposes
  has begun to see increased attention.
  Of particular interest in many environments is debloating commodity
  off-the-shelf (COTS) software, which is most commonly made
  available to end users as stripped binaries (i.e., neither source code
  nor symbol-table/debugging information is available).
  Toward this end, we created a system, called \toolname,
  that specializes stripped binaries.

  Many aspects of the debloating problem can be addressed via
  techniques from the literature on \emph{partial evaluation}.
  However, applying such techniques to real-world programs, particularly
  stripped binaries, involves non-trivial state-management manipulations
  that have never been addressed in a completely satisfactory manner in
  previous systems.
  In particular, a partial evaluator needs to be able to
  (i) save and restore arbitrary program states, and
  (ii) determine whether a program state is equal to one that arose earlier.
  Moreover, to specialize stripped binaries, the system must also be able
  to handle program states consisting of memory that is \emph{undifferentiated}
  beyond the standard coarse division into regions for the stack,
  the heap, and global data.

  This paper presents a new approach to state management in a program specializer.
  The technique has been incorporated into \toolname.
  Our experiments show that our solution to issue (i) significantly decreases the space
  required to represent program states, and our solution to issue (ii)
  drastically improves the time
  for producing a specialized program (as much as 13,000x speedup).

\Omit{
  To address (i), we present a mechanism for saving and restoring program states
  efficiently, which uses two OS-level mechanisms---copy-on-write (CoW) and process
  context-switching---to reduce time and space costs. For (ii), we present an
  application of Rabin's fingerprinting method, as a way to compute an
  efficient, collision-resistant, and incrementally updatable hash of a program
  state. Together, these mechanisms can work with arbitrary
  undifferentiated address spaces that cannot be reliably subdivided into
  high-level program variables at specialization time.

  Our experiments show that incremental hashing significantly improves the time
  for producing a specialized program (as much as 13,000x speedup), and
  that CoW significantly decreases the space required to represent
  program states.
}

\end{abstract}



\keywords{}

\maketitle


%
%
%

\section{Introduction}
\label{Se:Introduction}

Modern commodity off-the-shelf (COTS) software tends to provide large sets of
features to support the diverse use cases of their end-users. However,
individual users of many COTS programs might only use a single, fixed subset of
the available functionality. From such a user's perspective, unused
functionality constitutes ``bloat'' in terms of binary size, program
performance, and attack surface. A means of producing specialized versions of
programs that only include features relevant to a given use case would be a
useful tool for simplifying and hardening COTS software. In particular, given
certain configuration settings, a developer or administrator may wish to remove
features irrelevant to their particular configuration, thereby improving space
usage and performance, and reducing the program's attack surface.

Toward this end, we have created a system, called \toolname,
that specializes stripped binaries.
The premise behind our work is that many aspects of the ``debloating''
problem can be addressed via techniques from the literature on
\emph{partial evaluation} \cite{SCC:Futamura71,BOOK:JGS93}\nocite{HOSC:Futamura99}.
For instance, a partial evaluator $\PE$ takes as inputs
(i) a program $P$ (expressed in some language $L$);
(ii) a partition of $P$'s inputs into two sets, \textit{supplied}
and \textit{delayed} (for short, $S$ and $D$, respectively);\footnote{
  We find ``supplied'' and ``delayed'' to be more suggestive than the standard
  terms ``static'' and ``dynamic,'' respectively.
}
and
(iii) an assignment $\assignment{S}$ to the variables in $S$.\footnote{
  Here, the partition of $P$'s inputs is implicit in $\assignment{S}$.
}
As output, $\PE$ produces a \textit{residual program} $P_{\assignment{S}}$
that is specialized with respect to $\assignment{S}$.
More formally, we have
\begin{equation}
  \label{Eq:PartialEval}
  \llbracket \PE \rrbracket (P, \assignment{S}) = P_{\assignment{S}},
\end{equation}
where $\llbracket \cdot \rrbracket$ denotes the meaning function for
the language in which $\PE$ is written.
The requirement on residual program $P_{\assignment{S}}$ is that it
must obey the following equation:
\begin{equation}
  \label{Eq:OriginalToResidual}
  \llbracket P \rrbracket_L(\assignment{S \cup D}) = \llbracket P_{\assignment{S}} \rrbracket_L(\assignment{D}),
\end{equation}
where $\llbracket \cdot \rrbracket_L$ is the meaning function for $L$.
That is, $P_{\assignment{S}}$ with input $\assignment{D}$ produces the same output as $P$
with input $\assignment{S \cup D}$;
however, $P_{\assignment{S}}$ has fewer input arguments, and is
specialized with respect to the assignment $\assignment{S}$.

A partial evaluator may be able to identify parts of a program's control-flow graph (CFG)
that are unreachable given particular configuration settings,
and produce a residual program that does not contain the identified parts.
Moreover, code in the program that is dependent solely on the supplied inputs
can be executed by the partial evaluator, and elided from the resulting specialized
program.
In practice, these abilities allow a partial evaluator to perform a
multitude of optimizations, without the developer of the partial evaluator
needing to write explicit implementations of each optimization \cite{BOOK:JGS93}.
For example, a partial evaluator will perform removal of unreachable code
and constant folding, as well as more sophisticated optimizations,
such as loop-unrolling and function in-lining.
For debloating, a partial evaluator can (i) simplify code so that the resulting
program incorporates specific features based on particular configuration
parameters, and (ii) collapse abstraction layers in the original program via
function in-lining.

In some contexts---including in our work---an alternative formulation
of the above approach, based on the creation of \textit{generating
extensions}, is more desirable.
A generating extension can be thought of as a \emph{self-contained,
program-specific partial evaluator}.
A generating extension for $P$ and a specified set of supplied inputs
$S$ is a program $\GE_{P,S}$ that obeys the following equation:
\begin{equation}
  \label{Eq:GeDef}
  \llbracket \GE_{P,S} \rrbracket (\assignment{S}) = P_{\assignment{S}},
\end{equation}
where $P_{\assignment{S}}$ is the specialized residual program defined
previously, which obeys \eqref{OriginalToResidual}.
This approach to program specialization is enabled by a tool
called a \emph{generating-extension generator}: a program that
takes as input $P$ and $S$, and creates a generating extension
$\GE_{P,S}$.

The difference between the two approaches to program specialization
can be summarized as follows:
\begin{itemize}
  \item
    Applying a partial evaluator to program $P$ and partial state
    $\assignment{S}$ is similar to \emph{interpreting} $P$ on an input state, except
    that the output is a specialized program $P_{\assignment{S}}$.
  \item
    Applying a generating-extension generator to $P$ is similar to
    \emph{compiling} $P$, except that the outcome is a program, $\GE_{P,S}$,
    that, when executed on a partial state $\assignment{S}$, produces the
    specialized program $P_{\assignment{S}}$.
\end{itemize}

Generating extensions, and in particular machine-code generating
extensions, have the advantage that they can execute as native
programs; a semantic model of the target language is only needed to
\textit{produce} the generating extension.
At specialization time, no semantic information is needed (other than the
semantics built into the hardware platform on which a generating extension runs).

Moreover, a pre-made generating extension can be delivered to an end
user who wishes to specialize a program without needing to deliver
additional special-purpose tools for specializing programs.
\emph{For these reasons, we chose to work with generating extensions.}
 
Program specializers have been created for many different types of
languages, including imperative, functional, and logic-based,
both for source code and---less frequently---for machine code.
However, when creating a program-specialization tool for real-world programs,
one faces a multitude of problems.
In particular, a program-specialization tool must address two
state-management problems:
\begin{enumerate}
  \item
    \label{It:SavingAndRestoringStates}
    A program specializer needs to be able to save and restore program
    states efficiently.
  \item
    \label{It:IdentifyingStateRepetition}
    A program specializer uses a worklist-based algorithm that
    executes a program over partial program states (\sectrefs{HLL}{GEGenAlg}).
    To prevent redundant exploration of the program's state space,
    there needs to be an efficient means of determining whether
    a (partial) state has \emph{repeated}.
\end{enumerate}
Naive approaches to these issues are extremely costly (\sectref{ImplementationAndExperiments}):
\begin{itemize}
  \item
    A straightforward approach to \issueref{SavingAndRestoringStates} means
    copying the entire state for each save and restore operation.
  \item
    The need to test a new state against all states that have previously
    arisen (\issueref{IdentifyingStateRepetition}) suggests the use of hashing.
    However, resolving collisions requires the ability to compare two states for
    equality.
\end{itemize}
These state-management operations have never been addressed
in a completely satisfactory manner in prior work, and
the disadvantages of prior approaches become more
significant in the context of specializing stripped binaries.
For instance, programs often use \emph{linked data structures},
constructed using nodes allocated from the heap.
However, for a stripped binary, program states consist of memory that
is \emph{undifferentiated} beyond the coarse division into regions for the
stack, the heap, and global data.
Moreover, for a program specializer that runs natively, the states
that need to be captured and compared in
\issuerefs{SavingAndRestoringStates}{IdentifyingStateRepetition}
are \emph{native hardware states} (at the level of the instruction-set architecture).

In this paper, we describe a \emph{new technique for state management in a
  program specializer that runs natively}.
To demonstrate these techniques, we
implement a new program-specialization tool, \toolname, and present an evaluation of its
effectiveness. 
To address issues (\ref{It:SavingAndRestoringStates}) and
(\ref{It:IdentifyingStateRepetition}), our approach makes use of several
ideas known from the literature:
\begin{itemize}
  \item
    Using built-in OS process-creation and context-switching mechanisms for
    saving and restoring states \cite{exe06}.
  \item
    Using Rabin fingerprinting to create an incrementally updatable
     hash of a program's entire address space where there is an exponentially
     small probability of the hash of any two states colliding.\cite{nguyen08,mehler06}
  \item
    Exploiting hardware support for copy-on-write (CoW) memory
    management \cite{TODS:Lorie77} to identify changed memory regions,
    without the need to instrument arbitrary subject-program memory
    accesses or resorting to machine-code interpretation.
    Moreover, the use of CoW reduces physical memory pressure by
    sharing unchanged pages between multiple processes.
\end{itemize}
The contributions of our work are as follows:
\begin{enumerate}[label=\Alph*.]
  \item 
    Our main technical contributions are to state management in a program specializer
    that runs natively (\sectrefs{Overview}{StateManagement}).
    Unlike prior approaches used in program specializers, our state-management
    technique \emph{does not support a mechanism to resolve hash collisions}.
    Instead, by choosing appropriate values for parameters of the hashing scheme,
    the probability of a collision can be made arbitrarily small (in our
    case, $< 2^{-56}$), which allows
    us to forgo the conventional constraint that collisions be resolvable.
    By relaxing the collision-resolution constraint to a probabilistic guarantee, we obtain the following benefits:
    \begin{enumerate}[label=\arabic*.]
      \item
        With these technique, state equivalence can be checked in \emph{constant time}.
      \item
        This state-management technique handles program states over an address space
        divided into otherwise \emph{undifferentiated} stack, heap, and global regions.
        Fine-grained knowledge about variables and types is not required at specialization time.
        Nor is it necessary for the tool to have knowledge of the distinction
        between free storage and storage that is in use in the heap.
        \Omit{
        This approach is particularly beneficial for specializing stripped binaries,
        which lack symbol-table information.}
      \item
        Moreover, we are able to use a hashing technique that supports
        efficient incremental updating of hash values \cite{broder93,rabin81}.
    \end{enumerate}
  \item
    With these state-management techniques, we implemented \toolname, a new tool for
    specializing binaries.
    We present an evaluation of our technique's effectiveness in
    \sectref{ImplementationAndExperiments}.
\end{enumerate}
\Omit{To the best of our knowledge, our technique is a novel approach to state
management in a program specializer. } Our approach has several benefits:
\begin{itemize}
  \item
    It allowed us to create a program specializer that specializes machine code,
    runs natively, and can work without symbol-table information
    (\sectrefs{GEGenAlg}{StateManagement}).
  \item 
    The ability to perform O(1) state comparisons significantly improves
    specialization performance, compared to a naive approach (\sectref{ImplementationAndExperiments}).
  \item
    The use of CoW dramatically reduces memory usage, compared to using full-state copies
    (\sectref{ImplementationAndExperiments}).
  \end{itemize}

To make the paper self-contained, \sectref{HLL} presents a summary of
\emph{partial evaluation} and \emph{generating extensions}, using an
example to provide intuition.
\sectref{RelatedWork} discusses related work.
\sectref{Conclusion} concludes.



\section{A Pr\'ecis on Program Specialization}
\label{Se:HLL}
\Omit{Before discussing machine-code generating extensions, we describe partial
evaluation and subsequently generating extensions for C programs. In particular,
we describe a running example: the naive substring-matching procedure shown
\texttt{match} in \figref{HLStrMatch}(a). In \sectref{HLPE} we describe how a
partial evaluator specializes \texttt{match} on the pattern string. In
\sectref{GEOverview} we describe a C generating extension that performs the same
specialization. The construction of a machine-code implementation of the
generating extension is discussed in \sectref{GEGenAlg}.

We choose to discuss partial evaluation before generating extensions, as both
partial evaluators and generating extensions use the same worklist-based
state-space exploration strategy. Thus, it is easier to describe the state
exploration strategy without considering the additional concerns inherent in
constructing a generating extension, whether or not the generating extension is
constructed in a high-level language.
}

The purpose of this section is to provide background for readers
unfamiliar with partial evaluation and generating extensions, and to
help them understand how the material in
\sectseqref{Overview}{StateManagement} represents an advance over
previous work.
(Readers already familiar with these techniques may wish to peruse the
examples in this section and proceed to \sectref{Overview}.)
To aid understanding, relevant concepts are presented using
source-code examples, using the naive substring-matching procedure
\texttt{match} (\figref{HLStrMatch}(a)) as an example.

\sectref{HLPE} describes how a \emph{partial evaluator}
specializes \texttt{match} on the pattern string.
\sectref{GEOverview} describes a C \emph{generating extension} that
performs the same specialization.
In both approaches, there is a first phase of \textit{binding-time
analysis} (BTA) and a second worklist-driven \emph{specialization
phase} that produces the residual program.
Given the desired partition of the inputs into supplied and delayed
sets, BTA extends the partition to the program's variables at all
program points, identifying variable occurrences that can safely be
included in partial states.
The specialization phase traverses the subject program's CFG,
executing each basic block it encounters.
Moreover, the subject program is executed over \textit{partial
states}: states whose values can be safely computed when program
execution starts with an assignment to the supplied input variables.

\Omit{The construction of a generating extension for a machine-code version
of \texttt{match} is presented in \sectref{GEGenAlg}.}

\subsection{Overview of Partial Evaluation}
\label{Se:HLPE}
\begin{figure}
  \centering
{\footnotesize
  \begin{tabular}{@{\hspace{0ex}}c@{\hspace{0ex}}c@{\hspace{0ex}}}
    \begin{minipage}{.63\columnwidth}
\begin{alltt}
int match(char *p, \framebox{char *s}) \{
  while(\framebox{*s != 0})\{
    \framebox{char *s1 = s;} //block 2
    char *pat = p;
    while(1) \{
      if(*pat == 0)\framebox{return 1;} //block 3
      \framebox{if(*pat != *s1)} break; //block 4
      pat++; \framebox{s1++;} //block 5
    \}
    \framebox{s++;}
  \}
  \framebox{return 0;}
\}
\end{alltt}
    \end{minipage}
    &
    \begin{minipage}{.37\columnwidth}
  \begin{verbatim}
int match_s(char *s){
  while(*s != 0){
    char *s1 = s;
    if(*s1 == 'h'){
      s1++;
      if(*s1 == 'a'){
        s1++;
        if(*s1 == 't'){
          return 1;
        }
      }
    }
    s++;    
  }
  return 0;
}
\end{verbatim}
    \end{minipage}
    \\
    (a) & (b) 
  \end{tabular}
}
\caption{(a) String-matching program \texttt{match};
  (b) \texttt{match} partially evaluated on \texttt{p = "hat"}.
  \label{Fi:HLStrMatch}
}
\end{figure}

\newsavebox\phasedparsingone
\begin{lrbox}{\phasedparsingone}\begin{minipage}{.66\textwidth}%
\begin{alltt}
int match_ge(char *p)\{
  worklist_t L = empty_worklist();
  state_t successor_state;
  state_t cur_state;
  int cur_block;  
  worklist_enqueue(L, 1,init_state);
  printf("match_s(char *s)\{");
  while(!is_empty(L))\{  
    cur_block = get_worklist_head(L).block;
    cur_state = get_worklist_head(L).state;    
    remove_worklist_head(L);
    if(!previously_visited(cur_block, cur_state))\{
      if(cur_block == 1) handle_block_1(L, cur_state);
      if(cur_block == 2) handle_block_2(L, cur_state);
      //code elided
      if(cur_block == 8) handle_block_8(L, cur_state);
    \}
    printf("\}");
\}
\end{alltt}
  \end{minipage}%
\end{lrbox}

\newsavebox\phasedparsingtwo
\begin{lrbox}{\phasedparsingtwo}\begin{minipage}{.66\textwidth}%
\begin{alltt}
      void handle_block_3(worklist_t L, state_t S)\{
        pat = S.pat;
        printf("blk_3_%d:", S.id);
        successor_state = snapshot(pat);
        if(*pat == 0)\{         
          printf(" goto_7_%s", successor_state);
          worklist_enqueue(L, 7, successor_state);
        \}else\{
          printf(" goto_4_%s", successor_state);
          worklist_enqueue(L, 4, successor_state);
        \}
      \}
    \end{alltt}
 \end{minipage}%
\end{lrbox}

\newsavebox\phasedparsingthree
\begin{lrbox}{\phasedparsingthree}\begin{minipage}{.66\textwidth}%
\begin{alltt}
      void handle_block_4(worklist_t L, state_t S)\{
        pat = S.pat
        printf("blk_4_%d:", S.id);
        successor_state = snapshot(pat);
        \framebox{printf("if(%c != *s1)", *pat);}
        printf("  goto blk_2_%s", successor_state);
        printf("else goto blk_5_%s", successor_state);
        worklist_enqueue(L, 2, successor_state);
        worklist_enqueue(L, 5, successor_state);      
     \}
      void handle_block_5(worklist_t L, state_t S)\{
        pat = S.pat;
        printf("blk_5_%d:", S.id);
        pat++;
        \framebox{printf("s1++;");}
        successor_state = snapshot(pat);
        printf("goto blk_3_%s", successor_state);
        worklist_enqueue(L, 3, successor_state);
      \}
\end{alltt}
  \end{minipage}%
\end{lrbox}

\begin{figure*}
  \centering
  \begin{minipage}{.33\textwidth}%
    \centering%
    \resizebox{\textwidth}{!}{\usebox\phasedparsingone}%
  \end{minipage}%
  \begin{minipage}{.33\textwidth}%
    \centering%
    \resizebox{\textwidth}{!}{\usebox\phasedparsingtwo}%
  \end{minipage}%
  \begin{minipage}{.33\textwidth}%
    \centering%
    \resizebox{\textwidth}{!}{\usebox\phasedparsingthree}%
  \end{minipage}

  \captionsetup{width=0.8\linewidth}
  \caption{Generating extension for the naive string matcher from
    \figref{HLStrMatch}.
    \label{Fi:HLMatcherGE}
  }
\end{figure*}

The C procedure \texttt{match} in \figref{HLStrMatch}(a) is an implementation of
an $O(|s||p|)$ naive substring-matching algorithm. It returns 1 if and only if
the string pointed to by \texttt{s} contains the string \texttt{p} as a
substring. Note that \texttt{s} and \texttt{p} are presumed to point to valid C
strings, and thus \texttt{match} terminates whenever the null terminator (ASCII
0) for either string is encountered.

If we partially evaluate \texttt{match} with \texttt{p} pointing to the string
``hat'', we obtain the procedure shown in \figref{HLStrMatch}(b). In this
version, the inner loop has been unrolled, and all manipulations and uses of
\texttt{pat} and \texttt{p} have been eliminated: the characters in ``hat'' are
hard-coded into the tests in the specialized procedure. For this example,
\eqrefs{PartialEval}{OriginalToResidual} become
\begin{equation*}
  \llbracket \PE \rrbracket (\texttt{match}, [p \mapsto "hat"]) = \texttt{match}_{[p \mapsto "hat"]} = \texttt{match\_s},
\end{equation*}
\begin{equation*}
  \llbracket \texttt{match} \rrbracket_{\textrm{C}}([p \mapsto "hat"] \cup \assignment{D}) = \llbracket \texttt{match\_s} \rrbracket_{\textrm{C}}(\assignment{D}),
\end{equation*}
where $\llbracket \cdot \rrbracket_{\textrm{C}}$ denotes the meaning function
for C.

Partial evaluation can be implemented using a two-stage process, consisting of
BTA and the \textit{specialization phase}, which specializes the program by
executing over partial states \cite{BOOK:JGS93} (starting with an initial
partial state, such as $[p \mapsto "hat"]$).

There are many possible partitions that a BTA algorithm could produce.
A BTA algorithm is acceptable for our purposes as long as the partition that it
produces for each program point is \emph{congruent} \cite{BOOK:JGS93}.
Informally, congruence ensures that in every subject-program statement that
updates a supplied variable, the update to the supplied variable does not depend
on any delayed values. A partition of the variable occurrences at the different
program points of p into supplied and delayed sets ($V_s$ and $V_d$,
respectively) is congruent if at every statement $l$ in $P$ where a variable $v
\in V_s$ is updated, the new value of $v$ is computed solely from variables in
$V_s$.
Congruence is important because it ensures that the partial state
induced by the set of supplied inputs can always be safely updated.

A BTA algorithm can use \textit{forward slicing} \cite{ICSE:Weiser81,kn:HRB90}
to compute a congruent partition.
Given a set of variables $V$ and a set of program points $L$, forward slicing
computes the the set of program points that may be affected by the values of $V$
at points in $L$. For BTA, we compute the forward slice from the \textit{delayed
  inputs}. The boxed statements in \figref{HLStrMatch}(a) show the program
points included in the forward slice starting at formal parameter \texttt{s}. A
congruent partition of the program variable occurrences is implicit in the slice.
The forward slice contains all assignments to, and uses of, variable occurrences
that are transitively dependent on \texttt{s}, while the complement of the slice
contains all assignments to and uses of variable occurrences \textit{not}
dependent on \texttt{s}. Thus, to ensure that the specialization phase only
performs safe updates, it executes only the statements in the complement of the
slice. Moreover, slicing can be viewed as an extension of BTA results from
variable occurrences to statements: all statements dependent only on supplied
state are marked as supplied; the remainder are marked as delayed.

The specialization phase is essentially a kind of interpreter
that executes $P$ over partial states, producing a residual program $P'$.
The specializer interprets the CFG of the program, using a
partial state to track the values of the variable occurrences in the supplied
set. The interpretation is non-standard because at a condition classified as
delayed, such as the two boxed conditionals in \figref{HLStrMatch}(a), there are
\emph{two} successor basic blocks to interpret. A worklist is used to keep track
of basic blocks that still need to be processed.
Every basic block is interpreted linearly, statement-by-statement, and each
statement is evaluated in one of three ways. (1) All statements marked as ``supplied''  are
evaluated, and the partial state is updated accordingly. For example, 
the statement \texttt{pat++} will cause the value of \texttt{pat} in the
partial state to be incremented by 1. (2) Statements marked as ``delayed'' are not
evaluated, but are emitted to the residual program instead. For instance, the
single occurrence of ``\texttt{s1++}'' in the original \texttt{match} program is
emitted at two different times during the specialization of \texttt{match}. (3)
However, some statements marked as ``delayed'' cannot just be emitted as is;
if a delayed statement $s$ depends on the value of a supplied variable
$v$, the value of $v$ must be \textit{lifted} into the residual
program's state at $s$. Lifting can be performed by replacing every occurrence of $v$ in
the emitted statement with the current value of $v$. For example, lifting is
required for the \texttt{if} statement in the inner loop of \texttt{match}:
every emitted instance of the statement in \figref{HLStrMatch}(b) has
\texttt{*s1} replaced with a character from ``hat''. 

Unlike a standard interpreter, the specialization phase is prepared to handle
control flow governed by delayed state. Consider the \texttt{if} statement at
the end of the basic block marked as block 4 in \figref{HLStrMatch}(a). Due to
the comparison against the (delayed) string pointed to by \texttt{s1}, there is
not sufficient information in the partial state to determine which branch will
be taken. Consequently, the specializer must arrange to specialize the blocks at
\emph{both} successors.

In essence, the specializer needs to ``go both ways'' when encountering a branch
governed by delayed state. In practice, the specializer is generally implemented
as a worklist-based algorithm: basic blocks are specialized and residuated using
the approach described earlier; however, upon reaching a branch classified as
``delayed,'' the specializer records the current state, $\sigma$, and adds a
$(\sigma, l)$ pair to the worklist for every successor block $l$. The
specializer then removes an $(s,b)$ pair from the worklist, and executes basic
block $b$, starting with state $s$. Thus, at the basic-block level,
specialization is similar to execution, except that code can also be emitted; at
the end of a basic block, the specializer creates the appropriate
$(\text{partial-state},\text{basic-block})$ pair(s) for the block's successor(s), and inserts them
into the worklist.

The partial evaluation of \texttt{match} illustrates why a partial
evaluator needs to be able to check state equality efficiently.
Consider block 2, which contains the two assignments at the start of
the outer while loop, and ends with an unconditional branch into the
inner loop.
Every time block 2 is executed, \texttt{pat} is set to point to the start of
string \texttt{p}, and block 3 is enqueued. When block 3 is removed from the
worklist, the partial evaluator continues to unroll the inner loop.
Subsequently, the partial evaluator reaches the break statement following block
4, triggering a new partial evaluation of block 2: \texttt{pat} is reset, and
block 3 is again enqueued, ultimately leading to another identical unrolling of
the inner loop. Thus, a partial evaluator that \textit{always} enqueues the successor of
block 2, namely block 3, will never terminate.

To prevent this infinite unrolling, the partial evaluator must be able to detect
duplicate partial-state/block pairs. In particular, the first time we evaluate
block 2, we want to enqueue the pair $(\sigma,\, \text{block 3})$ consisting of
the state $\sigma$ where \texttt{p} is equal to \texttt{pat} and block 3.
Every subsequent time that a partial evaluation of block 2 is complete, we have
re-encountered the state-pair $(\sigma,\,\text{block 3})$. The partial evaluator
will not terminate unless it can determine that $(\sigma,\,\text{block 3})$ has 
repeated.

Thus, a worklist-based partial-evaluation algorithm requires two key
state-management features:
\begin{enumerate}
\item the ability to save and restore partial states,
\item the ability to efficiently check state equality
\end{enumerate}
When partial evaluation is performed on a program written in a type-safe
high-level language, both features can be implemented in a relatively
straightforward fashion. Assume that \texttt{match} is always called such that
the pointers \texttt{p} and \texttt{s} are guaranteed to reference the beginning
of valid C strings. In this case, the relevant state is the set of all supplied
variables and memory objects reachable from the supplied variables on the stack.
States can be saved, restored, and compared by traversing the graph of memory
objects induced by the reachability relation over the supplied state, in a
manner similar to the walk performed by a mark-and-sweep garbage collector.

\Omit{Such an approach can be implemented efficiently by use of an abstract
  datatype of partial states for which saving/restoring states and identifying
  state repetition can be performed with low time and space overhead. In
  particular, the components of (partial) states can be hash-consed
  \cite{UT-ISL-TR-74-03:Goto74} so that a unique representative---i.e., a
  canonical address---is maintained for each partial state. A set of the
  addresses of the unique representatives is then maintained, with hashing used
  to assist membership testing (and collision resolution performed merely by
  comparing two addresses).}

In \sectref{StateManagement}, we describe an alternative method for state
management that is more suitable for generating extensions, particularly
machine-code generating extensions (\sectrefs{GEOverview}{GEGenAlg}).

\subsection{Overview of Generating Extensions}
\label{Se:GEOverview}
An alternative approach to program specialization can be implemented via a
\textit{generating-extension generator}.
\Omit{Whereas a partial evaluator
is analogous to an interpreter, a generating-extension generator is
analogous to a compiler.}
A generating-extension generator $\mathbf{GeGen}$ takes as input a
program $P$ and the BTA results, and produces a \textit{generating
extension}:
\begin{equation*}
\llbracket \mathbf{GeGen} \rrbracket (P,S) = \GE_{P,S}
\end{equation*}
where the generating extension $\GE_{P,S}$ produces a residual program:
\begin{equation*}
\llbracket \GE_{P,S} \rrbracket(\assignment{S}) = P_{\assignment{S}}
\end{equation*}
such that $P_{\assignment{S}}$ satisfies \eqref{OriginalToResidual}.
A generating extension has two key advantages over a partial evaluator:
\begin{enumerate}
\item It can be implemented as a program that executes natively in the target
  language, without interpretation. A semantic model of the target language is
  only needed to \textit{construct} the generating extension.
\item The structure of a generating extension reflects the basic-block structure of the subject program.
\end{enumerate}

Structurally, a generating extension can be thought of as the original subject
program, with the partial-evaluation code ``compiled in.''
This intermingled structure can be structured in such a way that generating extensions can be
algorithmically produced basic-block-by-basic-block.
Each basic block in the subject program has an associated basic-block procedure in the generating
extension that updates the partial state of the subject program and, 
generates residual code.
After these actions are completed, the block yields control to the compiled-in state-management logic.
This structure was used by Andersen \cite{AndersenDissertation} to automatically produce generating
extensions for C programs, and we use a similar approach to structure our
generating extensions for machine code (but with different state-management
mechanisms that are described in \sectref{StateManagement}).

For example, \figref{HLMatcherGE} is an Andersen-style C generating extension for
procedure \texttt{match} from \figref{HLStrMatch}(a).
Consider procedure \texttt{match\_ge} in \figref{HLMatcherGE}:
\texttt{match\_ge} repeatedly dequeues a $(\text{partial-state},\text{basic-block})$ pair $(\sigma, b)$ from the
worklist until the worklist is empty.
If $(\sigma, b)$ has not been visited yet, block $b$ is
evaluated on $\sigma$ through a ``basic-block-procedure.''

The \texttt{state\_t} struct and \texttt{snapshot} procedure are used to save
and restore states. The \texttt{state\_t} struct has a member for every variable
in the partial state (in this case, just \texttt{pat}), and \texttt{snapshot}
stores the current values of the variables in the struct. To illustrate the
structure of ``basic-block-procedures,'' we consider three basic blocks from the
inner loop of \figref{HLStrMatch}(a): blocks 3 and 4, which end with the two if
statements in the inner loop, and block 5, which increments the two string
pointers.

The structure of each basic-block-procedure reflects the structure of the basic
blocks in the original program. The statements dependent only on the contents of
the string pointed to by \texttt{s} (the boxed statements in
\figref{HLStrMatch}(a) and \figref{HLMatcherGE}, excluding the if statement) are
merely quoted and printed verbatim. Conversely, the statements dependent only on
the supplied value \texttt{p} (the unboxed statements in \figref{HLStrMatch}(a))
are evaluated during the execution of the generating extension. The statement
\texttt{if(*pat != *s1) break;} depends on both \texttt{pat} and the delayed
string \texttt{*s1}. Variable \texttt{pat} must be lifted: the statement is
printed as written, except that supplied variable \texttt{pat} is replaced with
its current value.
\Omit{
\begin{figure}
  \includegraphics[width=3.0in]{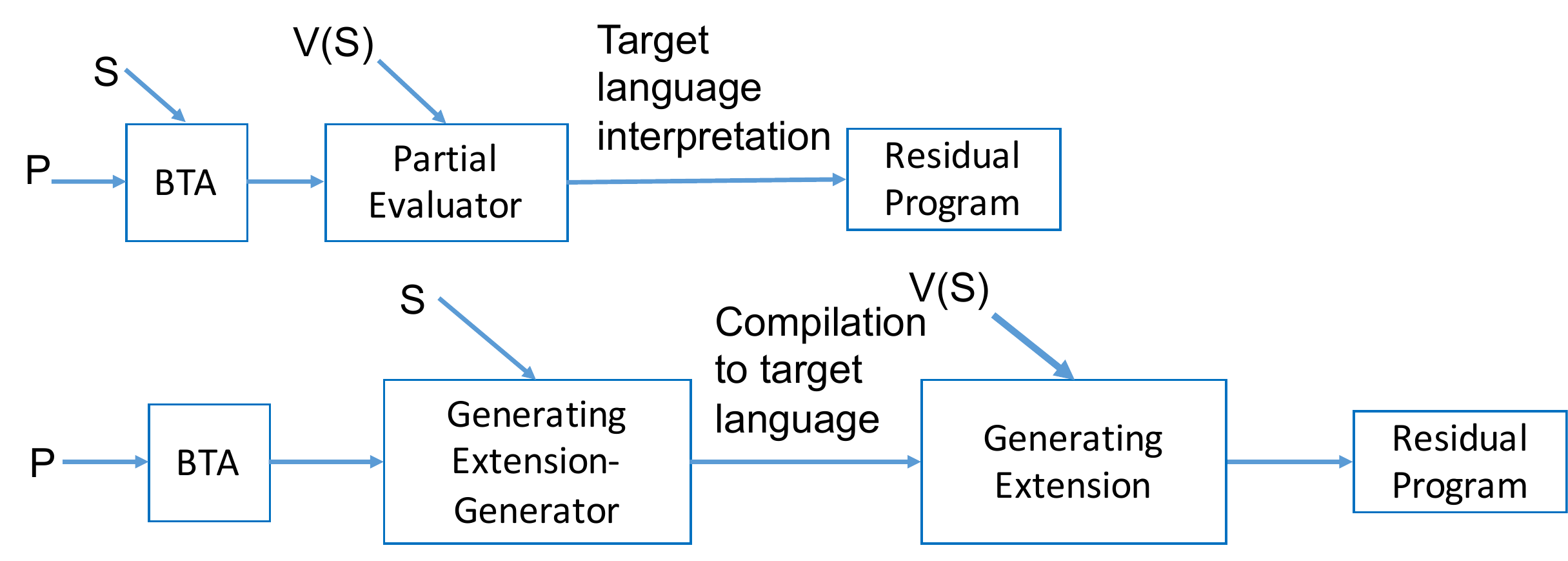}
  \caption{Partial evaluators are analogous to interpreters, while generating-extension generators are analogous to compilers.\label{Fi:GePictoral}}
\end{figure}
}

\Omit{
\begin{figure}
  \begin{small}
\begin{verbatim}

\end{verbatim}
    \end{small}
  
\end{figure}
}

The correspondence between the generating extension and the subject program
makes it straightforward to create a generating extension algorithmically.
While the execution of the generating extension is now worklist-driven and
non-standard, the basic block-level structure still reflects that of the
original subject program. Moreover, the generating extension itself need not
include a C interpreter; a special-purpose analysis tool is only necessary to
construct the generating extension itself.

Internally, however, the generating extension must still handle the
two\Omit{ key} state-management issues described in \sectref{Introduction}:
(1) saving and restoring partial states, and (2) checking state equality.
At the end of each basic-block procedure,
the generating extension takes a snapshot of its own state,
emits code to transfer control to its successor block(s),
and inserts each successor block into the worklist.
Finally, control returns to the top of the loop in \texttt{match\_ge},
and another state/block pair is dequeued if the worklist is not empty.

\Omit{Additionally, the execution of generated code is simpler to
  reason about than the execution of a general-purpose partial evaluator.}

\section{Overview}
\label{Se:Overview}

The best prior solution to the state-management problem has been to
take advantage of the fact that a partial evaluator is
similar to a language interpreter
\cite{BOOK:JGS93}---except that a partial evaluator operates on
partial states, and an interpreter operates on full states.

One can design an abstract datatype of partial states for which
saving/restoring states and identifying state repetition can be
performed with low time and space overhead.  In particular, the
components of (partial) states can be hash-consed
\cite{UT-ISL-TR-74-03:Goto74} so that a unique representative---i.e.,
a canonical address---is maintained for each partial state.\footnote{
  More precisely, to support the unique-representative property, one would make use of
  applicative maps (see \cite[\S6.3]{toplas:RTD83} and \cite{popl:Myers84}),
  hash-consing, and a hash table to detect duplicates.
  (The hash-code
  would be based on the contents of the map's entries, rather than the
  structure of the tree that represents the map.)  } A set of the
addresses of the unique representatives is then maintained, with
hashing used to assist membership testing (and collision resolution
performed by comparing addresses).

  \Omit{
For the work described in this paper, such an approach was
unsatisfactory due to one of the design decisions that we made. We
wanted to avoid having to interpret machine code, and thus chose the
generating-extension-generator approach for specializing binaries.
}

For our work on specializing binaries, such an approach was unsatisfactory.
We chose the generating-extension \Omit{generator }approach, because it creates program
specializers that end-users can use without learning sophisticated
program-analysis tools.
We wanted to avoid (i) packaging a full-featured
interpreter for x86 with the specializers or (ii) instrumenting every load and
store in the subject binary.
Consequently, we did not have the option of
implementing memory as an explicit data structure that can be readily swapped
to save and restore states---which raises the following
question:
\begin{equation*}
\begin{array}{|p{.95\columnwidth}|}
\hline
  \textrm{How can \issuerefs{SavingAndRestoringStates}{IdentifyingStateRepetition}
  from \sectref{Introduction} be handled efficiently in a generating extension $\GE_{P,S}$ that runs natively?}\\
\hline
\end{array}
\end{equation*}

To address \issueref{SavingAndRestoringStates}, we use
two OS-level mechanisms---copy-on-write (CoW) and process
context-switching---to create an efficient mechanism for
\textit{state-snapshotting and restoration}.
(See \sectref{snapshots}.)
However, the main element that allowed us to devise a solution is that
we changed the requirements associated with
\issueref{IdentifyingStateRepetition} slightly.
In particular, \emph{we do not insist that there be a mechanism to
resolve collisions}, as long as we have control over parameters that
ensure that \emph{the probability of a collision \textbf{ever} arising
is below a value of our choosing}.
In other words, we allow the use of a \emph{collision-resistant} hash.
Moreover, the hash function is \emph{incrementally updatable}:
as execution of $\GE_{P,S}$ mutates one (partial) state $\sigma_1$ to
another (partial) state $\sigma_2$, the hash value for $\sigma_2$ can
be computed efficiently by updating the hash value of $\sigma_1$.
(See \sectref{StateHashing}.)

In our implementation, by using $128$-bit hash values, we ensure that when the
program uses $\le 2^{35}$ bits of memory and visits $\le
1\textrm{,}000\textrm{,}000$ unique states, the probability of the hash value of
any visited state colliding with the hash value of any other visited state is
less than $2^{-56}$. Thus, although it is possible for our tool to produce an
incorrect residual program due to a hash-value collision, the chances of that
happening are negligible.\footnote{ Put another way, it is as if we had arranged
  for a year to have the right number of days so that, with any randomly chosen
  group of $1\textrm{,}000\textrm{,}000$ people, the chances of winning a
  birthday-paradox bet was less than $2^{-56}$. } In our implementation,
incremental updating of hash values occurs at page granularity.

Our experimental results show that these mechanisms are critical to
the practical tractability of machine-code generating extensions. Most
significantly, if Rabin fingerprinting or some other $O(1)$ state-comparison
mechanism is not used, the amount of time to
produce a residual program scales quadratically with the number of
partial states seen. Even on simple examples, fingerprinting yields
several orders-of-magnitude of improvement in execution times; one
simple test case required over 12 hours without fingerprinting,
while requiring only 3 seconds with fingerprinting.

Using CoW produces a large improvement in the amount of space
needed to represent partial states; while thousands of pages of memory
are required to represent visited states without CoW, at most several
hundred are required when CoW is used in our
experiments. Moreover, when fingerprinting is used, CoW yields an
additional two-to-six-fold reduction in the time required to produce a
residual program.

\OnlySubmission{(An extended version of the paper has been submitted as supplementary material.)}


\section{Constructing Machine-Code Generating Extensions}
\label{Se:GEGenAlg}

In this section, we explain the technique for creating generating extensions
used in our tool \toolname.
\toolname takes a program and BTA results as input, and transforms each basic
block of the program into a self-contained unit that
(i) executes the basic block,
(ii) updates the partial state that depends on supplied input,
(iii) produces a specialized basic block,
(iv) snapshots the partial state, and
(v) yields control to a controller process.
The BTA results determine the transformation of individual instructions
performed in step (iii).

\toolname relies on the implementation of BTA from WiPER \cite{SR:Wiper}.
WiPER invokes CodeSurfer/x86 \cite{CC:BGRT05}, which incorporates
a number of algorithms for static analysis of machine code \cite{TOPLAS:BR10}
to build a dependence graph that supports machine-code slicing \cite{MCSlice}.
As in \sectref{HLPE}, BTA is performed by slicing forward from
the delayed inputs, marking all program points in the slice as
delayed, and all points outside the slice as supplied.

To identify lifted values, reaching-definition analysis is performed
for the operands of each instruction $I$ in the delayed set.
Any static instructions that define an operand used by $I$ must have
associated code to lift the value of the operand.

\begin{figure}
{\footnotesize
\begin{alltt}
L3:  \doublebox{mov dl, [ebx]}    -- dereference pat
     cmp dl, 0          -- check first if condition
     jz L7              -- if(*pat == 0) return 1
L4:  \framebox{mov cl, [eax]}    -- dereference s1
     \framebox{cmp cl, dl}       -- check second if condition
     \framebox{jne L2}           -- if(*pat != *s1) break;
L5:  \framebox{incr eax}         -- s1++
     incr ebx           -- pat++
     jmp L3             -- while(1)
\end{alltt}
}
  \caption{Naive string matcher's inner loop body. Boxed instructions are
    delayed, double-boxed instructions have their destination operands lifted,
    and the remainder are supplied.\label{Fi:InnerBody}}
\end{figure}

In \figref{InnerBody}, BTA results are illustrated for the code that
implements the innermost loop of \texttt{match} from \figref{HLStrMatch}.\footnote{The 32-bit
  Intel x86 instruction set (also called IA32) has six 32-bit general-purpose
  registers (\asm{eax}, \asm{ebx}, \asm{ecx}, \asm{edx}, \asm{esi}, and
  \asm{edi}), plus two additional registers: \asm{ebp}, the frame pointer, and
  \asm{esp}, the stack pointer. In Intel assembly syntax, which is used in the
  examples in this paper, the movement of data is from right to left (e.g.,
  $\asm{mov eax,ecx}$ sets the value of \asm{eax} to the value of \asm{ecx}).
  Arithmetic and logical instructions are primarily operand instructions
  (e.g., $\asm{add eax,ecx}$ performs $\asm{eax := eax + ecx}$). An operand in
  square brackets denotes a dereference (e.g., if \asm{a} is a local variable
  stored at offset -16, $\asm{mov [ebp-16],ecx}$ performs $\asm{a := ecx}$).
  Branching is carried out according to the values of condition codes
  (``flags'') set by an earlier instruction. For instance, to branch to \asm{L1}
  when \asm{eax} and \asm{ebx} are equal, one performs \asm{cmp eax,ebx}, which
  sets \asm{ZF} (the zero flag) to 1 iff $\asm{eax} - \asm{ebx} = 0$. At a
  subsequent jump instruction $\asm{jz L1}$, control is transferred to \asm{L1}
  if $\asm{ZF} = 1$; otherwise, control falls through.}
Register \texttt{eax} contains the address of the current offset in the string that
is being searched, and \texttt{ebx} contains the address of the current offset in the
pattern to be matched.
The registers \texttt{cl} and \texttt{dl} contain the
current characters in the string and pattern, respectively.
Basic blocks \texttt{L3}, \texttt{L4}, and \texttt{L5} correspond to the inner loop
blocks in \figref{HLStrMatch}.
\texttt{L7} is reached only if a match is found.
\texttt{L2} is the target of the inner loop's break statement, which
starts another iteration of the outer loop.

In lifted instruction $\texttt{mov dl,\,[ebx]}$, register \texttt{dl} must be
lifted, because \texttt{cmp cl,\,dl} in block \texttt{L4} compares
the supplied pattern character in \texttt{dl} to a character
from the delayed string. 

Given the partitioning of instructions in \figref{InnerBody}, the
generating-extension-generator emits x86 code augmented with
\textit{pseudo-instructions} that expand to sequences of x86 instructions.
Their actions emit code, control the flow of computation, and manage
partial states.

\begin{figure}
{\footnotesize
    \begin{tabular}{@{\hspace{0ex}}c@{\hspace{2.5ex}}c@{\hspace{0ex}}}
      \begin{minipage}{1.6in}
          \begin{alltt}
L3: EmitSpecLabel(L3)
    \doublebox{mov dl, [ebx]}
    \doublebox{Lift(dl)}
    cmp dl, 0
    MakeSnapshot
    EmitSpecJmp("jz", L7, L4)
    CondEnqueue("jz", L7, L4)
    Yield
  \end{alltt}
\end{minipage}
      &
        \begin{minipage}{2.0in}
            \begin{alltt}
L4: EmitSpecLabel(L4)
    \framebox{Emit("mov cl, [eax]")}    
    \framebox{Emit("cmp cl, dl")}
    MakeSnapshot
    \framebox{EmitDynJmp("jne", L2, L5)}
    Enqueue(L2, L5)
    Yield
L5: EmitSpecLabel(L5)
    \framebox{Emit("incr eax")}
    incr ebx
    MakeSnapshot
    Enqueue(L3) 
    EmitJmp(L3)
    Yield
  \end{alltt}
\end{minipage}
\end{tabular}
}
\vspace{-2.0ex}
    \caption{\label{Fi:InnerBodyGE}The machine-code generating extension produced
      for the code in \figref{InnerBody}.
      The three blocks are the machine-code analogs of the
      \texttt{handle\_block} functions in \figref{HLMatcherGE}.}
\end{figure}

\figref{InnerBodyGE} illustrates the machine-code generating extension
produced from \texttt{match}.
In this presentation, we treat the pseudo-instruction actions as black boxes.
As before, non-branch instructions classified as supplied are
executed, updating the partial state.
Conversely non-branch instructions classified as delayed are emitted
verbatim in a manner analogous to the printf calls in
\figref{HLStrMatch}.

The lifted instruction $I =\;$\texttt{mov dl,\,[ebx]} is executed, just like
a supplied instruction. After $I$ is executed, \texttt{Lift} emits code
that sets the value of \texttt{dl} in the residual program to the value that
\texttt{dl} holds immediately after the execution of $I$.

\texttt{MakeSnapshot} records a snapshot of the current partial state of the
subject program, using the technique described in \sectref{snapshots}.
  
Both the \texttt{Enqueue} and \texttt{CondEnqueue} actions place
$(\text{partial-state}, \text{basic-block})$ pairs into the worklist. In both
cases, the state used for every enqueued pair is the state recorded by the most
recent call to \texttt{MakeSnapshot}, and a pair is only enqueued if it has
not been enqueued before.

\texttt{Enqueue} can be used to enqueue a single successor, in the case of
unconditional jumps, or multiple successors, in the case of jumps controlled
by delayed state. \texttt{CondEnqueue} is used for conditional jumps governed
by supplied state; its action is to enqueue the successor block determined by the
supplied state and the condition of the jump instruction.

The actions of \texttt{EmitJmp}, \texttt{EmitSpecJump}, and \texttt{EmitDynJump}
are to emit jumps to specialized versions of a block's successors. If the
original jump is conditional and governed by supplied state, we use
\texttt{EmitSpecJmp} to emit an unconditional jump to the specialized version of
whichever successor block is chosen by the jump's condition. If the original
jump is conditional and governed by delayed state, \texttt{EmitDynJmp} is used
to emit a conditional jump targeting the specialized versions of the block's two
possible successors. If the jump is unconditional, we use \texttt{EmitJmp} to
emit an unconditional jump to the specialized version of the only successor.

The action of \texttt{EmitSpecLabel} is to emit a label unique to
the current $(\sigma, b)$ pair being invoked.

The action of \texttt{Yield} is to emit code that yields control to the controller
process. The controller process is a simple piece of code that removes a pair
$(\sigma, b)$ from the worklist, restores state $\sigma$, and resumes execution at
block $b$.

\paragraph{Pointers.}
The implementation of the \texttt{Lift} macro needs to correctly handle pointers
to stack and heap objects.
A pointer to a heap or stack
object during specialization time may not be a valid pointer to the object in
the residual program. \Omit{Despite its limitations in other aspects, }CodeSurfer/x86's
VSA implementation reliably identifies whether a memory location or
register holds a pointer to the stack or heap at a given program point in all of the
programs tested. By combining this information with a special-purpose
implementation of \texttt{malloc} used only in the generating extension,
concrete pointers into memory objects can be converted into relocatable offsets.



\section{OS-Assisted State Management}
\label{Se:StateManagement}

\Omit{
\Omit{If we wish to produce a natively executing generating extension
that does not have a massive amount of overhead from, e.g.,
instrumenting loads and stores, we cannot use the
state-management techniques available to a partial evaluator,
such as the hash-consing-based approach mentioned in
\sectrefs{Introduction}{HLPE}.}
This section describes how to implement efficient state-management
primitives to support a machine-code generating extension,
as described in \sectref{GEGenAlg}.
}

\subsection{Implementation of Snapshots}
\label{Se:snapshots}
\Omit{
Because a partial evaluator saves and restores states at the end of every basic
block, it is critical for these operations to be implemented in an efficient
fashion. While saving and restoring CPU state is a straightforward operation,
saving and restoring memory state is potentially expensive, both from the
standpoint of storage required to represent states, as well as the amount of
time required to save and restore states. Because the generating extension
implements subject-program logic natively in assembly code, we do not have the
option of implementing memory as an explicit data structure that can be readily
swapped to save and restore memory configurations.
}
Because a generating extension saves and restores states at the end of every basic
block, it is critical for these operations to be implemented efficiently.
While saving and restoring CPU state is a straightforward operation,
saving and restoring memory state is potentially expensive, both from the
standpoint of storage required to represent states, as well as the amount of
time required to save and restore states.

We use different OS processes to represent different state snapshots, and the
\texttt{fork()} system call to generate a new snapshot of interest. Thus, a set
of snapshots---in our case, the elements of the generating extension's
worklist---can be represented by a set of process IDs. A partial state can be
restored efficiently merely by performing a process switch.

\Omit{We still need to address the issue of supporting the ability to
determine efficiently whether two partial states are identical.  Note
that comparing process IDs is not enough: there can be two identical
partial states that have different process IDs.  Our solution to this
problem is described in \sectref{StateHashing}.}

The \texttt{fork()} system call is implemented in a relatively time
and space-efficient fashion through the use of a policy known as
\textit{copy-on-write} (CoW). Through the use of hardware-supported
virtual memory, logical addresses used by a process are decoupled from
their physical addresses in memory. The address space of a process is
broken up into fixed-size pages (typically 4096 bytes), each of which
can be mapped to an arbitrary physical memory page. This decoupling
lets processes share physical pages, enabling CoW.

When a process $P$ calls \texttt{fork()}, a new process $P'$ is
created with register and memory contents identical to $P$, with the
exception of \texttt{eax}, which contains the return value of
\texttt{fork()}. Rather than allocating new physical pages for $P'$,
every page $G_{(P',i)}$ in $P'$ is mapped to the same physical page
$H$ as the corresponding $G_{(P,i)}$ in $P$. However, the
virtual-to-physical mappings for $P'$ are flagged as CoW
using hardware support; when $P'$ writes for the first time to a page $G_{(P',i)}$
inherited from $P$, a hardware fault occurs, the
changed version of the page is allocated its own page $H'$ in physical
memory, and the hardware state is updated so that $G_{(P',i)}$ is 
mapped to $H'$.

\begin{SCfigure}
  \includegraphics[width=2.3in]{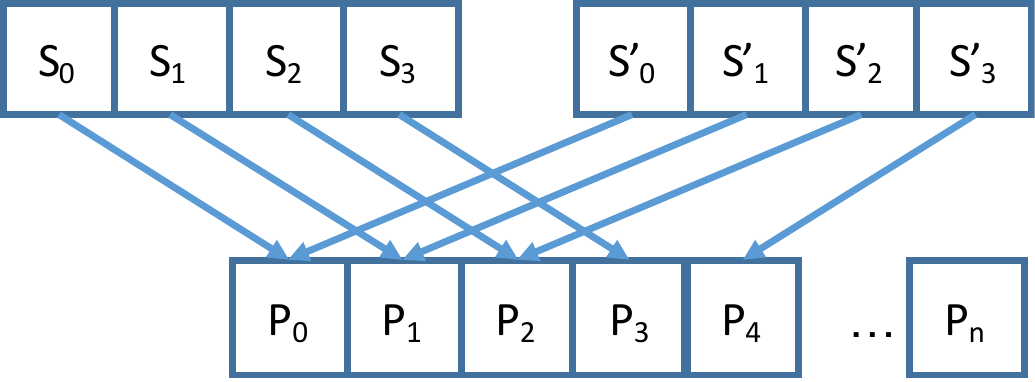}
  \caption{\label{Fi:CoW}States $\sigma$ and $\sigma'$ after a CoW fault at $P_3$.}
\end{SCfigure}

Consider the two processes $\sigma$ and $\sigma'$ in \figref{CoW},
where $\sigma'$ is the result of forking $\sigma$ and executing some code that
modifies the fourth page of the virtual address space. Each
process has a four-page virtual address space, backed by an $n$-page
physical memory. In this case, the first three virtual pages of both
processes map to the same physical pages, while the fourth page of
each process maps to different physical pages. 

Our generating extensions exploit CoW to implement the end-of-block worklist
update. Each partial state referred to in \sectref{HLL} is a separate Linux
process. An additional ``controller'' process oversees the partial evaluator's
worklist of $(\text{partial-state},\text{basic-block})$ pairs. The controller process serves as
a dispatcher for the specialization phase, and the worklist of unprocessed
$(\text{partial-state},\text{basic-block})$ pairs is implemented merely as a set of process IDs.
Every time the controller process selects a $(\text{partial-state},\text{basic-block})$ pair
$(\sigma, b)$ from the worklist, it signals the process $P_{\sigma}$ that
represents $\sigma$, and $P_{\sigma}$ begins executing. $P_{\sigma}$
immediately calls \texttt{fork()}, creating a child process $P'$. Initially,
the logical address space of $P'$ contains $\sigma$, and its program counter
is set to $b$. After $P'$ finishes executing $b$, its logical address space
contains $\sigma'$. However, physically only the pages that \emph{changed}
during the execution of $b$ on $\sigma$ are specific to $P'$; the rest are
shared with process $P_\sigma$. Because all the memory bookkeeping is
implicitly taken care of by the OS and the hardware, the only data state that
is stored and manipulated by the controller process is the ID that the OS
assigns to each process.

Thus, executing a basic block $b$ on a given state $\sigma$ incurs a cost that
is linear in the number of memory-writing instructions in $b$, because each
such instruction is executed only once. Moreover, the cost of switching
between processes is constant: a switch consists of a system call and the
update of several fixed-size hardware registers.

\subsection{State Hashing}
\label{Se:StateHashing}

For a generating extension to avoid traversing previously seen computation paths,
it needs an efficient way to determine whether a given state has been seen before.
We desire a state-management scheme that possesses five properties:
\begin{enumerate}[label=(\roman*)]
  \item
    \label{It:NoFalseNegatives}
    Given a state $\sigma$ that has been seen before,
    whenever we encounter $\sigma$ again, the procedure must recognize $\sigma$ as
    a previously visited state, with no false negatives.
  \item
    \label{It:SpaceAndTimeEfficient}
    The procedure must be space- and time-efficient.
    We would like to store at most several hundred bits of information per state
    visited. We would also like $O(1)$ state-equality checks.
  \item
    \label{It:LowFalsePositiveRate}
    The false-positive rate must be kept acceptably low.
  \item
    \label{It:IncrementallyUpdatable}
    We would like to be able to efficiently update the value
    that characterizes a visited state.
  \item
    \label{It:SpecialCaseForZeroPages}
    If the initial state is a multi-gigabyte address space,\Omit{(for
    example a 4-gigabyte (i.e., $2^{35}$-bit) address space)} computing
    the initial value that characterizes the state should require a
    constant amount of computation.
\end{enumerate}
These criteria naturally suggest a solution based on hashing.
Note that \itemref{LowFalsePositiveRate} deviates from the conventional
approach to state management in program specialization, which is that there
should be \emph{no} false positives.
As discussed in \sectref{Introduction} and explained below, although
it is possible for our tool to produce an incorrect residual program
due to a hash-value collision, the chances of that happening are negligible.
Our choice to work with this relaxed requirement was motivated by the
fact that our generating extensions work with native hardware states.

To produce efficient generating extensions,
\itemref{IncrementallyUpdatable} is especially important. When computing the
post-state hash after executing a single block, we wish to perform an amount of
computation proportional to the number of changes to the pre-state made during
the block's execution. By exploiting properties of
CoW, it is possible to satisfy \itemref{IncrementallyUpdatable} with
an appropriate choice of hash algorithm:
\begin{itemize}
\item When a basic block $b$ is executed by the specializer, the first write to
  a page in the execution of $b$ induces a CoW fault. Given a log of
  all CoW faults that occur during the execution of the basic block,
  we can compute the \emph{changes} between the pre-state $\sigma$ and
  the post-state $\sigma'$. Implementing this log was
  straightforward: we added (i) a small amount of instrumentation code to the
  Linux kernel's page-fault handler, and (ii) a small amount of extra
  state to every process structure in the kernel.
  \item
    We need an incrementally updatable hashing algorithm---one that
    lets us efficiently incorporate differing pages into
    the pre-state hash-code, without additional computation beyond
    processing of the data of changed pages in the pre- and post-states.
    Rabin's fingerprinting scheme\footnote{Though Rabin fingerprinting is most
      well-known for its use as a sliding-window hash, it can also be used for
      incremental hashing \cite{rabin81}.} satisfies this criterion \cite{broder93,rabin81}.
\end{itemize}
Formally, given the pre-state $\sigma$ and its associated hash
$H(\sigma)$, and the contents of the changed pages,
$P_{\textit{pre}} = \{P_1, ... P_n\}$, $P_{\textit{post}} = \{P_1',...P_n'\}$, from the
pre- and post- states, we compute the post-state hash 
using only the pre-state hash $H$, and the contents
of $P_{\textit{pre}}$, and $P_{\textit{post}}$:
\begin{equation*}
  \label{Eq:IncrUpdate}
  H(\sigma') = H_{\textit{incr}}(\sigma, P_1, P_1', ..., P_n, P'_n)
 \end{equation*}

 Given a bit-string $\sigma = (s_0, s_1 ... s_{m-1})$, representing the contents of a
 program's address space, we wish to compute a hash $H(\sigma)$. To do so, the
 fingerprinting algorithm treats $\sigma$ as a polynomial $\sigma(t) = s_0* + s_1*t^1 +
 ... + s_{m - 1}*t^{m-1} $ of degree $m - 1$ with coefficients over $Z_2$.
 The fingerprinting scheme selects an irreducible polynomial $P(t)$ (i.e.,
   $P(t)$ is only divisible by 1 and itself) of degree $k$, again with
 coefficients from $Z_2$. Given $P(t)$, the fingerprint $H(\sigma)$ is defined
 as
\begin{equation*}
  \label{Eq:fpDef}
  H(\sigma) = \sigma(t)\mod P(t)
\end{equation*}
\Omit{That is, $H(\sigma)$ is the bit-string of coefficients of $H(\sigma)\mod P(t)$} 
The choice of the degree $k$ of the irreducible polynomial $P$ allows us
to choose the size of the hash, and thereby tune the collision
probability relative to a definition of a ``reasonable'' execution of
a program specializer. For our purposes, we are assuming that the
partial state of the subject program will use a $2^{35}$-bit address
space and will visit 1 million unique states. Given these assumptions,
simple counting arguments outlined in \cite{broder93} and
\cite{rabin81} show that for a 128-bit hash code (i.e., $k=127$), the probability
of there being \emph{any} collision among the 1 million hash-codes
is less than $2^{-56}$.

Note that the evaluation of the polynomial at a
value of $t$ plays no part in fingerprinting: we merely use the algebraic
properties of the polynomials themselves. Specifically, polynomials with
coefficients over $Z_2$ have several properties convenient for the
implementation of an incrementally updatable hash:
\begin{enumerate}
\item
  \label{It:AdditionIsXor}
  The addition operation $+$ for such polynomials is addition mod two with no
  carry---i.e., bitwise exclusive-or. Consequently, subtraction for polynomials
  with coefficients over $Z_2$ is simply addition. These properties let us
  treat the contents of memory as $\sigma(t)$, thus incurring no additional space
  overhead.
\item
  \label{It:MultiplicationIsShift}
  Multiplication by $t^i$ can be implemented as an $i$-bit shift.
\item
  \label{It:Linearity}
  Fingerprinting is linear:
  \begin{equation*}
    H(A + B) = H(A) + H(B)
\end{equation*}
\item
  \label{It:Decomposibility}
  The fingerprint of the product of $t^i$ and a polynomial $\sigma(t)$ can be
  computed via
  \begin{equation*}
    H(t^i * \sigma(t)) = H(H(t^i) * H(\sigma(t)))
  \end{equation*}
\end{enumerate}
Given property
(\ref{It:AdditionIsXor}) of polynomials over $Z_2$, in what follows,
we will use $\sigma$ to denote both the bit-string representation of $\sigma$ and
the polynomial $\sigma(t)$; the intended use will be clear from context.

Consider the pre and post-state in \figref{CoW}, where virtual page 3
maps to physical page 3 in the pre-state and physical page 4 in the post-state.
Properties (\ref{It:AdditionIsXor})-(\ref{It:Decomposibility}) admit a
simple update procedure: 
\begin{equation*}
  H(\sigma') = H(\sigma) + H(\boxed{P_3}) + H(\boxed{P_4})
\end{equation*}

From (\ref{It:AdditionIsXor}), it follows that given a change to the
$i^{\textit{th}}$ page in pre-state $\sigma$, the post-state $\sigma'$ can be
derived by subtracting off the terms representing the contents of the
$i^{\textit{th}}$ page in $\sigma$ and adding on the terms corresponding to the
post-state version of the page in $\sigma'$. By coupling this observation with
properties (\ref{It:MultiplicationIsShift}), (\ref{It:Linearity}), and
(\ref{It:Decomposibility}), it can be shown that the $H(\sigma')$ can be
directly computed from $H(\sigma)$ using only the contents of the
$i^{\textit{th}}$ page in $\sigma$ and $\sigma'$, avoiding the need to examine
all of $\sigma'$ to compute its hash value.

In particular, let $\pagelength$ be the page size in bits supported by
the OS (here $4096 * 8 = 2^{15}$ bits).  In addition, let $\sigma_{a,b} = s_a +
s_{a+1}t + ... + s_b*t^{b - a}$
denote the bit-string containing the bits of the substring of $\sigma$
starting at $a$ and ending at $b$, inclusively, for both $a$
and $b$.

Then, from properties (\ref{It:AdditionIsXor}) and (\ref{It:Linearity}), we have
\begin{equation*}
H(\sigma') = H(\sigma) + H(t^{i*w}*\sigma_{i*w,(i+1)*w-1}) + H(t^{i*w}*\sigma'_{i,(i+1)*w-1})
\end{equation*}
and by property (\ref{It:Linearity})
\begin{equation*}
H(t^{i*\pagelength}*\sigma_{i*\pagelength,(i+1)*\pagelength-1}) =
H(H(t^{i*\pagelength})*H(\sigma_{i*\pagelength,(i+1)*\pagelength-1}))
\end{equation*}
For a fixed page size of, e.g., 4096 bits, the only non-constant-time
computation is $H(t^{i*\pagelength}) = t^{i*\pagelength} \mod P$, which can be
computed in time $\log_2(i*\pagelength)$ using 
modular-exponentiation-via-squaring.
Because the maximum amount of addressable memory is
bounded on x86 CPUs, $\log_2(i+w)$ is effectively a small constant in practice.

The number of pages that must be hashed in order to compute the post-state hash
is $O(m)$, where $m$ is the number of unique pages written during the execution
of the basic block.
In the common case, $m$ at most $O(n)$, where $n$ is the number of instructions
in the basic block.
Thus, for the common case, hashing induces a constant overhead on the amount of
computation performed by a basic block.
The only exceptions are special x86 opcodes, such as those that use the
\texttt{rep} prefix; these instructions essentially implement loops that perform
memory writes repeatedly, until some condition is met. These instructions are
often used for, e.g., string operations.
In the programs we examined, the use of \texttt{rep}-prefixed instructions to write large stretches
of memory is uncommon; we did not encounter any cases where \texttt{rep} was
used to write
regions larger than a page.
Additionally, in our semantic model we consider \texttt{rep}-prefixed
instructions to be loops, and we treat the individual prefix-free version of the
instruction as a basic block.

In addition to efficient incremental updates, this hash technique also handles
new and empty pages efficiently.
Any new physical memory added to a program's address
space is zeroed out by the OS for security reasons; thus, when a program's
address space grows, it will contain zeroes.
It is clear from the properties of reduction modulo a polynomial that the hash
of a zero page is zero; thus, no additional computation needs to be performed to
incorporate new pages into the hash of a program state.

\Omit{
This approach differs from the most common uses of Rabin
fingerprinting in the literature. Most commonly, Rabin fingerprints are
used to incrementally update a sliding hash; given a bit-string $\sigma$,
$|\sigma| = m$, a window of size $n < m$ slides over $\sigma$, and the algebraic
properties of the polynomials are exploited to implement the update
operation for an efficient sliding-hash algorithm.  However, Rabin
also uses the polynomial-based approach to create a method of hashing
bit-strings that admits in-place updates, which is what we need for
our purposes.}
\Omit{We now illustrate a specific example of the incremental hashing.
Consider the states $\sigma$ and $\sigma'$ in \figref{CoW}.
Focusing on the change from $\sigma$ to $\sigma'$, we see that only the changes to the
fourth page need to be incorporated into the post-state hash $H'$.

As illustrated in \figref{Update}, Rabin's scheme takes the contents of
pages $\sigma_3$ and $\sigma'_3$ (the contents of physical pages $P_3$ and
$P_4$), hashes the two pages individually, and folds them into the
updated hash.

\begin{figure}
  \includegraphics[width=2.8in]{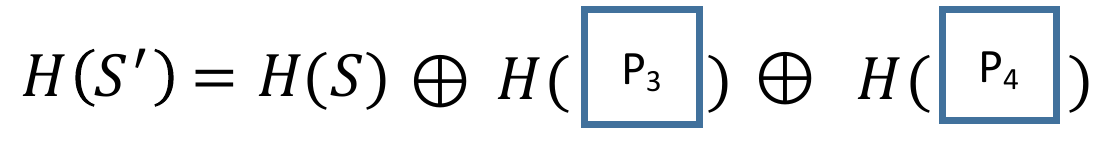}
  \caption{\label{Fi:Update}Computing the updated hash using the old hash-code,
  the old page contents, and the new page contents.}
\end{figure}

In particular, let $\pagelength$ be the page size in bits supported by
the OS (here $4096 * 8 = 2^{15}$ bits).  In addition, let $\sigma_{a,b} = s_a +
s_{a+1}t + ... + s_b*t^{b - a}$
denote the bit-string containing the bits of the substring of $\sigma$
starting at $a$ and ending at $b$, inclusively, for indices $a$
and $b$. The contents of $\sigma'_3$ in the pre-state polynomial can be
represented as
$s_{\pagelength * 3}t^{\pagelength * 3} + ...  + s_{\pagelength * 4 -
  1}t^{\pagelength * 4 - 1}$. We can factor out the low-order term,
$t^{\pagelength * 3}$, obtaining
$t^{\pagelength*3}(s_{w*3} + ... + s_{w*4 - 1}t^{\pagelength -
  1})$. Thus, the $i^{\textit{th}}$ page in the process can be treated as
$t^{\pagelength*i}*\sigma_{\pagelength,\pagelength-1}$. Thus, for this example,
$\sigma = \sum_{i = 0}^{3}t^{\pagelength*i}\sigma_{\pagelength*i,\pagelength*(i+1)-1}$. It

Hence, $\sigma' = \sigma - t^{\pagelength*3}*\sigma_{\pagelength*3,\pagelength*4-1} +
t^{\pagelength*3}*\sigma'_{\pagelength*3,\pagelength*4-1}$. Thus, $H(\sigma') = H(\sigma -
t^{\pagelength*3}*\sigma_{\pagelength*3,\pagelength*4-1} +
t^{\pagelength*3}*\sigma'_{\pagelength*3,\pagelength*4-1})$. However, the linearity
of $H$ (\eqref{linearity}), and the equivalence of $+$ and $-$
(\eqref{AdditionIsXor}) yield $H(\sigma') = H(\sigma) +
H(t^{\pagelength*3}*\sigma_{\pagelength*3,\pagelength*4-1}) +
H(t^{\pagelength*3}*\sigma_{\pagelength*3,\pagelength*4-1})$. It is clear that the
algebraic reasoning in this example generalizes to any arbitrary page $i$:
\begin{equation*}
H(\sigma') = H(\sigma) + H(t^{i*w}*\sigma_{i*w,(i+1)*w-1}) + H(t^{i*w}*\sigma'_{i,(i+1)*w-1})
\end{equation*}
Then, we can implement $H_{\textit{incr}}(\sigma,P_1,P_1',...P_n,P_n')$ in a straightforward
fashion. The pre- and post-state contents of an updated page, $P_i$, are simply
hashed and xor-ed with the pre-state hash.}



\section{Implementation and Experiments}
\label{Se:ImplementationAndExperiments}

\paragraph{Implementation.}
\label{Se:Implementation}

The work to create \toolname was reduced by adopting the BTA
implementation from the WiPER partial evaluator \cite{SR:Wiper},
which uses the slicing facilities provided by CodeSurfer/x86 \cite{CC:BGRT05}.
After BTA is performed on a program, \toolname traverses the program's
CFG, and emits the generating extension using the macros described in
\sectref{GEGenAlg}.
The generating extension is then assembled:
the generating extension is placed as inlined assembly inside a small
C++ wrapper, so the ``assembler'' is actually \texttt{g++}.
The generating-extension binary can then be given values for supplied
inputs, and run on a version of the Linux 4.4.14 kernel modified to
track CoW faults.
The final residual program is then assembled---this time using inlined
assembly inside a small C wrapper, so the assembler is \texttt{gcc}.
(The use of \texttt{g++} and \texttt{gcc} for assembly is an
implementation expedient.)

\Omit{\paragraph{Pragmatics.}
Two subject applications, \texttt{gnu-wc} and \texttt{gnu-env} are
modified versions of the corresponding GNU coreutils programs. The
applications were modified, with certain procedures replaced with
hand-written stubs. These changes were to accommodate limitations of
the current version of CodeSurfer/x86's VSA implementation and
library models, which disrupted BTA. These issues, however are not
intrinsic to our BTA strategy, and can be rectified in a future
version of CodeSurfer/x86.

The third subject program, \texttt{printf}, calls into a simple printf
library \cite{simplePrintf}. As with the coreutils programs, the printf library
was modified to address precision issues with VSA.
}

\paragraph{Experimental Questions.}
\label{Se:ExperimentalQuestions}

Our experiments were designed to answer the following questions:
\begin{enumerate}
\item
  \label{Qu:SpaceImprovement}
   What are the individual improvements to memory usage contributed by CoW and fingerprinting?
\item
  \label{Qu:TimeImprovement}
  What are the individual improvements to the time needed to emit a residual
  program contributed by CoW and fingerprinting?
\item
  \label{Qu:ResidSpeedup}
  Compared to the original subject program, how much does specialization
  speed up execution?
\end{enumerate}

\subsection{Specialization Performance}
\label{Se:GEPerf}

\begin{figure}
  \footnotesize
  \begin{tabular}{@{\hspace{0.0ex}}||@{\hspace{0.5ex}}l@{\hspace{0.5ex}}|@{\hspace{0.5ex}}l@{\hspace{0.5ex}}|@{\hspace{0.5ex}}l@{\hspace{0.35ex}}||@{\hspace{0.0ex}}}
    \hhline{|t:===:t|}
    Application & Description & Static Input \\
    \hhline{|:===:|}
    power       & Computes $x^n$ & $n = 100$ \\
    \hhline{||---||}
    dotproduct  & Computes the dot product of & $n = 100$, and coefficients \\
                & two $n$-dimensional vectors & of first vector \\
    \hhline{||---||}
    interpreter & Interpreter for the minimalist & an input program \\
                & language ``Brainf*ck''         & \\
    \hhline{||---||}
    filter      & Applies $m \times m$ convolution & $m = 3$, $n = 3$, and elements \\
                & filter on an image of size $n \times n$ & of the filter \\
    \hhline{||---||}
    sha1        & Computes the sha1 digest of a & $n = 1024$ and the contents \\
                & message of size $n$ bits      & of the first $512$ bits \\
    \hhline{||---||}
    matcher     & A naive substring-matching & the target substring \\
                & algorithm                  & \\
    \hhline{||---||}
    stack       & A program that writes every & $n$ \\
                & stack page $n$ times        & \\
    \hhline{|b:===:b|}
  \end{tabular}
 \caption{\label{Fi:subjectprogs}Microbenchmarks used in the evaluation.}
\end{figure}

\paragraph{Experimental Setup.}
\Omit{
\begin{figure}
  \begin{tabular}{@{\hspace{0.0ex}}||@{\hspace{0.5ex}}l@{\hspace{0.5ex}}||@{\hspace{0.5ex}}l@{\hspace{0.5ex}}|@{\hspace{0.5ex}}l@{\hspace{0.5ex}}|@{\hspace{0.5ex}}l@{\hspace{0.5ex}}||@{\hspace{0.0ex}}}
    \hhline{|t:====:t|}
                                 &                                  & \multicolumn{1}{@{\hspace{0.5ex}}c@{\hspace{0ex}}}{Supplied} & \multicolumn{1}{@{\hspace{-0.55ex}}|@{\hspace{0.5ex}}c@{\hspace{0.5ex}}||@{\hspace{0ex}}}{Delayed} \\
    \multicolumn{1}{@{\hspace{0.0ex}}||@{\hspace{0ex}}c@{\hspace{0ex}}}{App.} & \multicolumn{1}{@{\hspace{-1.05ex}}||@{\hspace{0.5ex}}c@{\hspace{0ex}}}{Description} & \multicolumn{1}{@{\hspace{-0.55ex}}|@{\hspace{0.5ex}}c@{\hspace{0ex}}}{input}    & \multicolumn{1}{@{\hspace{-0.55ex}}|@{\hspace{0.5ex}}c@{\hspace{0.5ex}}||@{\hspace{0ex}}}{input} \\
    \hhline{||====||}
    gnu-wc  & counts lines, chars,  & which quantities & stdin \\
            & or words in stdin     & to count         & \\
    \hhline{||----||}
    gnu-env & executes program       & assignment to    & program \\
            & with specified        & env. variables   & to invoke \\
            & environment           &                  &  \\
    \hhline{||----||}
    printf  & program that calls a  & format string    & remaining \\
            & simple printf library &                  & arguments \\
    \hhline{|b:====:b|}
  \end{tabular}
  \vspace{-1.5ex}
  \caption{The subject programs}
  \label{Ta:SubjectProgs}
\end{figure}
}
We evaluated \toolname using the binaries of seven microbenchmarks---listed in \figref{subjectprogs}---and,
as ``real-world'' examples, three command-line binaries:
two GNU coretuils programs, and one program that makes use of \texttt{printf}.
Five of the microbenchmarks were previously used to evaluate WiPER \cite{SR:Wiper}.\Omit{\footnote{
  In the WiPER paper, experiments were performed with three additional programs
  to test WiPER's ability to extract a component from a binary, which is a capability that
  \toolname does not support.
}}
The sixth, \texttt{matcher}, is the naive string matcher given in \figref{HLStrMatch}.
The seventh, \texttt{stack}, is designed to stress test the fingerprinting technique.\footnote{
  The ten benchmarks are provided as supplementary material.
}
\begin{itemize}
  \item
    \texttt{gnu-wc} counts lines, chars, or words in \texttt{stdin}. The supplied input
    specifies which quantities are counted; the delayed input is \texttt{stdin}.
  \item
    \texttt{gnu-env} runs a program with a specified assignment to
    environment variables. The supplied input is the assignment to environment
    variables; the delayed input is the program to invoke.
  \item
    \texttt{printf} is a program that calls into a simple printf library. The
    supplied input is a format string; the delayed input is the remaining arguments. 
\end{itemize}
These programs present a reasonable cross section of real-world
specialization tasks.
\texttt{gnu-wc} represents a \emph{feature-removal task}, in
which a single mode of operation is chosen out of a set of potential
modes, while \texttt{printf} is a fairly representative \emph{layer-collapsing and
loop-unrolling task}, in which a library call is in-lined into a program.
The third program, \texttt{gnu-env}, features aspects of both tasks, because the core
environment-update loop is unrolled, and features corresponding to unused
command-line flags are excised.\footnote{
  The reason we used only three real-world programs in our study
  was because of limitations of CodeSurfer/x86
  \cite{TOPLAS:BR10,CC:BGRT05}, which \toolname uses to implement
  BTA.
  \Omit{
  Because CodeSurfer/x86's analysis results are based on an overapproximation
  of the set of states that can actually arise during execution,
  the dependence graph constructed by CodeSurfer/x86 often contains
  spurious dependence edges---i.e., the graph contains an overapproximation
  of the actual dependences that exist among program elements.
  When BTA is performed on such a graph, some variables and instructions
  are identified as being delayed that one would like to be identified as supplied.
  The consequence is that the resulting generating extension cannot
  do much to specialize the subject program.

  \hspace*{1.5ex}}
  Fortunately, in many circumstances, the subject program
  can be adapted to overcome the limitations, e.g., by manually
  unrolling a loop. 
  However, the effort required to \Omit{examine the dependence graph of each
  program, determine the cause of analysis imprecision, }identify
  appropriate rewritings to overcome current limitations of the static
  analyses in CodeSurfer/x86, as well as to model calls to library
  functions, limited the number of real-world programs that we were
  able to use for our study.
}

For \texttt{gnu-wc}, specializing with respect to the supplied input
selects one of three main application loops, each of which is optimized for a
different counting task. The generating extension elides the other two loops.

In the case of \texttt{printf}, the specialization unrolls the format string,
eliminating run-time parsing and logic for unused format specifiers.
Similarly, in \texttt{gnu-env}, the argument-parsing loop is unrolled, emitting a 
program that runs a program in a pre-defined environment. 

To evaluate \questionrefs{SpaceImprovement}{TimeImprovement}, we
implemented \toolname so that CoW and state fingerprinting can be
independently disabled in generating extensions, yielding four
possible execution modes (see \figref{Results}).

To simulate disabling of CoW, we added a mechanism to force the copy of an entire
process address space.
When CoW is ``disabled,'' we dirty each page without altering
the state by (i) writing a single byte to each page in the address space, and
then (ii) reverting the page back to its original state.
These actions force every page \Omit{in the process }to incur a CoW
fault, causing the OS to create a copy of every page in the address space.
This approach provides an upper bound on the time required because,
by forcing the CoW mechanism to make the copy, a page
fault must be handled by the kernel for every page, adding some
overhead.
We chose to estimate the cost in this way because our generating
extensions are inherently multi-process: each process holds a single
state.
Implementing a true CoW-free approach would have required modifying
the OS to eliminate CoW, which seemed unwarranted, given that the
technique is not likely to be competitive.

To disable fingerprinting, we implemented an alternative version of the
generating extension's state-comparison and worklist-management algorithm.
Without fingerprinting, the only way to compare the states of two processes is
to do a direct comparison of process memory. Moreover, we no longer have a
convenient means of indexing into a table of previously seen states.
Consequently, the state manager must retain a process for every state previously
seen, and must compare every newly created process state with every retained
state, comparing full address spaces. In contrast, in the fingerprint-based
approach, we only need to store the 128-bit fingerprint; any process that does
not have outstanding worklist entries can be garbage-collected\Omit{, releasing any
pages not shared with descendant processes}.

To measure memory usage, the generating extension tracks the number of pages in use across
all processes in the generating extension.  
Because all processes must be retained when fingerprinting is
disabled, determining the memory usage across all processes is
straightforward: it is the sum of all live pages across
all processes.
When fingerprinting is used, memory usage is the maximum number of live pages at
any given point in the program's execution.
To evaluate the execution time of a generating extension, we time its
end-to-end execution, from the beginning of the first basic
block to the end of the last basic block. 

We allowed the generating extensions to run end-to-end for the
``real-world'' examples. However, for the microbenchmarks, we added a time-out
after 90 minutes of specialization.


\begin{figure}
  \centering \resizebox{1.01\columnwidth}{!}{
    \begin{tabular}{@{\hspace{0.0ex}}||@{\hspace{0.25ex}}l@{\hspace{0.25ex}}|l@{\hspace{0.7ex}}||c|c|c|c||c|c||@{\hspace{0.0ex}}}
      \hhline{~~|t:======:t|}
      \multicolumn{2}{c||}{} & \multicolumn{4}{c||}{Generating-extension performance} & \multicolumn{2}{c||}{Execution time} \\
      \hhline{~~||~~~~--||}
      \multicolumn{2}{c||}{} & \multicolumn{4}{c||}{ [CoW,Fingerprint]} & orig. & resid. \\
      \hhline{~~||----||~~||}
      \multicolumn{2}{c||}{} & [no,no] & [yes,no] & [no,yes] & [yes,yes] & prog. & prog. \\
      \hhline{|t:==#====#=|=:|}
      printf & time & 68m 23s & 66m 11s & 6.138s & .744s & 90.6 $\pm$ 5.1 $\mu$s & 77.7 $\pm$ 4.8 $\mu$s\\
      \hhline{||~-||----||-|-||}
      client & pages & 240577 & 48 & 12774 & 6 & --- & ---\\
      \hhline{|:=======|=:|}
      gnu-wc & time & 45m 36s & 50m 11s & 2.755s & 1.190s & 283 $\pm$ 7.8 $\mu$s & 106 $\pm$ 5.1 $\mu$s\\
      \hhline{||~-||----||-|-||}
                             & pages & 146901 & 46 & 2129 & 9 & --- & --- \\
      \hhline{|:=======|=:|}
      gnu-env & time & 13h 12m & 12h 26m & 15.692s & 3.332s & 36.5 $\pm$ .1 $\mu$s & 31.0 $\pm$ .1 $\mu$s\\
      \hhline{||~-||----||-|-||}
                             & pages & 958050 & 129 & 2129 & 2 & --- & ---\\
      \hhline{|:=======|=:|}
      power & time & 74m 39s  & 64m 36s  & 2.241s & .679s &  3.3 $\pm$ 0 $\mu$s& .6 $\pm$ 0 $\mu$s\\
      \hhline{||~-||----||-|-||}
                             & pages & 221416 & 102 & 2129 & 1 & --- & ---\\
      \hhline{|:=======|=:|}
      dotprod. & time & >90m & >90m & 11.366s & 2.364s & 3.7 $\pm$ 0 $\mu$s & .8 $\pm$ 0 $\mu$s\\
      \hhline{||~-||----||-|-||}
                             & pages & --- & --- & 2129 & 1 & --- & ---\\
      \hhline{|:=======|=:|}
      interp. & time & >90m & >90m & 13.638 & 6.186s & 35.9 $\pm$ .2 $\mu$s& 36.2 $\pm$ .1 $\mu$s\\
      \hhline{||~-||----||-|-||}
                             & pages & --- & --- & 2129 & 1 & --- & ---\\
      \hhline{|:=======|=:|}
      filter & time & >90m & >90m & 16.391 & 6.370 & 4.1 $\pm$ 0 $\mu$s & .6 $\pm$ 0 $\mu$s\\
      \hhline{||~-||----||-|-||}
                             & pages & --- & --- & 4258 & 2 & --- & ---\\
      \hhline{|:=======|=:|}
      sha1 & time & >90 m & >90 m& 24.223 & 11.783s & 4.6 $\pm$ 0 $\mu$s & 3.3 $\pm$ 0 $\mu$s\\
      \hhline{||~-||----||-|-||}
                             & pages & --- & --- & 2129 & 2 & --- & ---\\
      \hhline{|:=======|=:|}
      matcher & time & 28m 30s & 26m 13s &  1.652 & .839 & .9 $\pm$ 0 $\mu$s & .1 $\pm$ 0 $\mu$s\\
      \hhline{||~-||----||-|-||}
                             & pages & 185223 & 22 & 31935 & 3 & --- & ---\\
      \hhline{|:=======|=:|}
      stack & time & 3m 43s & 4m 36s & 1m 34s & 1m 26s &  5,533 $\pm$ 26.0 $\mu$s & .1 $\pm$ 0 $\mu$s\\
      \hhline{||~-||----||-|-||}
                             & pages & 195900 & 195900 & 1959 & 1959 & --- & ---\\
      \hhline{|b:========:b|}
    \end{tabular}
  } \vspace{-1.5ex}
  \caption{Run times and space usage for each generating extension, with and
    without CoW/fingerprinting. Run times for original and residual programs are
    also included, with 95\% confidence intervals (``---'' means ``not measured.'')}
  \label{Fi:Results}
\end{figure}

\begin{figure}
  \centering 
    \begin{tabular}{||c||c|c||}
      \hhline{|t:===:t|}
      program & original & residual\\
      \hhline{|:=#==:|}
      printf & 754 & 1038\\
      \hhline{||-||--||}
      gnu-wc & 1929 & 775 \\                       
      \hhline{||-||--||}
      gnu-env & 1820 & 1123 \\ 
      \hhline{||-||--||}
      power & 30 & 323 \\ 
      \hhline{||-||--||}
      dotprod. & 307 & 1123 \\ 
      \hhline{||-||--||}
      interp. & 146 & 558 \\ 
      \hhline{||-||--||}
      filter & 287 & 1207 \\ 
      \hhline{||-||--||}
      sha1 & 332 & 2823 \\ 
      \hhline{||-||--||}
      matcher & 34 & 410 \\ 
      \hhline{||-||--||}
      stack & 3930 & 1 \\ 
      \hhline{|b:===:b|}
    \end{tabular} 
  \caption{Instruction counts for original and residual programs.}
  \label{Fi:CodeSizeResults}
\end{figure}
\paragraph{Results.}
The experimental results for \questionrefs{SpaceImprovement}{TimeImprovement}
are presented in \figref{Results}. With respect to
\questionref{SpaceImprovement}, both fingerprinting and CoW play a significant
role in reducing memory usage.
Using CoW, however, yields the most significant
reduction for every application, except \texttt{stack}.
This improvement is due to the fact that for all ten applications, the
instructions that are evaluated during generating-extension execution
perform the majority of their writes within a single stack page.
Even when fingerprinting is not used, CoW ensures that the number of pages
needed to retain all previously visited states is small, roughly the
number of basic blocks that executed at least one memory write.

Regarding \questionref{TimeImprovement}, fingerprinting plays the most
significant role in reducing execution time. This result is unsurprising,
because the amount of time needed to identify whether a state has been
previously visited without using fingerprinting scales linearly with the number
of states previously visited. Thus, the execution time scales quadratically with
the number of states.

Stress-test \texttt{stack($n$)} performs a set of writes that causes
the generating extension to recompute each stack-page fingerprint $n$ times.
Still, the benefits of $O(1)$ lookup outweigh the cost of
repeatedly fingerprinting every stack page.

Using CoW also improves the execution times of generating extensions; the
improvement is most pronounced in the case where fingerprinting is also used.
When fingerprinting is used, the overhead of copying an entire process begins to
dominate the execution time of the generating extension.

For the \texttt{gnu-wc} and \texttt{stack} generating extensions without
fingerprinting, the execution time with CoW enabled was greater than when CoW
was disabled.
We do not have a full explanation, but we believe that the extra cost is due
to the cost of collecting memory-usage data.
When we measure memory usage with CoW enabled, we track every process
currently using a given page.
For certain workloads, especially when fingerprinting is not
used---and thus page mappings are retained for every state
visited---the cost of maintaining this data structure may become
relatively large.

\subsection{Speedup of Specialized Programs}
\label{Se:ResidProgPerf}

  To evaluate experimental \questionref{ResidSpeedup}, we timed the end-to-end
  execution time of each program on an input, collecting the 10\% trimmed mean of 1001
  executions: i.e., for the original and residual version of each program, we ran the
  program 1001 times, and discarded the 100 shortest and 100 longest execution
  times.
  
  To time the programs, we instrumented the beginning and end of \texttt{main}
  in each program with calls to a \texttt{rdtscp}-based timer. By doing so we
  avoid recording the noise induced by the initial context-switching, loading,
  and execution of the program. The hardware-counter-based \texttt{rdtscp} counter provides ~$40$-clock-cycle resolution.
  
  For \questionref{ResidSpeedup}, results are presented in \figref{Results}.
  Specialization produced a speedup in all but one program.
  Because \texttt{stack} has no meaningful delayed actions, it has a
  5500x speedup, due to the elision of thousands of memory writes.
  The most significant speedup in a specialized program with non-trivial delayed functionality was for \texttt{matcher}, which was 9x faster.
  This improvement can be attributed to the specialization of the inner loop of the program,
  which elides all the memory loads for the target string.
  In particular, this change speeds up the common case in which a
  character in the string being searched does not match the first
  character of the target string. \texttt{filter} yields the second most
  significant speedup, being 6.8x faster.
  The specialization of filter significantly optimizes the inner loop of the
  image filtering procedure, eliminating the \texttt{if} statement that selects
  which image filter is applied to each pixel, as well as inlining loads from
  lookup tables that encode properties of the selected filter algorithm.

  \texttt{power} and \texttt{dotproduct} benefit significantly from the unrolling
  of their main loops; the elimination of the branch condition at the loop head yield a 5.5x speedup and
  a 4.6x speedup, respectively. 

  The specialized version of \texttt{gnu-wc} has a speedup of 2.7x 
  The specialization of \texttt{gnu-wc} elides the argument-parsing loop, as well as
  setup code that (i) sets locale information and (ii) obtains system-dependent
  configuration information.

  \texttt{gnu-env}, \texttt{printf}, enjoy more modest speedups, roughly 16-18\%.
  Most of the speedup is due to the unrolling of the core loop in each program:
  for format-string parsing in \texttt{printf}, and the argument-parsing and
  environment-setup loops in \texttt{gnu-env}.

  \texttt{sha1} obtains a 1.4x speedup from the elision of loads inside the main loop, along with the
  elision of the initial code that initializes the supplied data.

  However, \texttt{interpreter} experiences a slight slowdown (< 1.1x) ,  possibly due due to the effects of
  aggressive unrolling on cache performance.
\section{Related Work}
\label{Se:RelatedWork}

\Omit{Related work falls into two categories:
(i) work related to partial evaluation of machine code, and
(ii) incremental hashing.
\Omit{(We are not aware of work similar to our use of process forking
and CoW together to implement the kind of fine-grained snapshots
that our generating extensions use.)}
}

\subsubsubsection{Specialization of Machine Code.}
\Omit{ Run-time code generation is a generating-extension-like approach to
  program specialization that produces machine code on-the-fly during program
  execution. Unlike our approach to machine-code specialization, which operates
  on stripped \textit{binaries} without source code or symbol-table information, run-time
  code generation performs BTA as part of compilation. In systems that support
  run-time code generation, programs must be developed in a specific high-level
  language with specialization in mind. The user provides annotations for a
  high-level-language program, which assist or replace binding-time analysis,
  and the compiler produces a machine-code generating extension for each
  annotated region.

For example, in the Fox system \cite{leone96,PLDI:LL96}, type-level
information in ML source code is exploited to produce run-time
machine-code generators to be invoked by other parts of the ML program. 
The arguments of a procedure are divided into two tuples:
the first component contains the supplied arguments;
the second component contains the delayed arguments.
For example, the parameters to an ML implementation of the three-argument
function $\texttt{p}$ from \figref{HLStrMatch}(c) would be defined in the following
fashion:
\[
  \texttt{fun p (s) (d1, d2)}
\]

The run-time code-generation systems avoid the state-management
problems discussed in \sectref{StateManagement} by either using
semantic information available at compile time from the high-level code,
as with Fox \cite{leone96,PLDI:LL96} and Lancet \cite{rompf14}, or by supporting code-generation
strategies that make no attempt to conserve resources by
discovering repeated states, as in \cite{engler96pe}. In Fox and
Lancet, the exploration of program points at redundant states is
prevented by ensuring that the code generator for an annotated
function $P$ is called at most once with any given set of actuals. In
the case of \cite{engler96pe}, the annotation schemes are essentially
domain-specific languages for writing run-time code generators. The
user-implemented code generators are roughly analogous to Lisp macros,
except that unlike Lisp macros, code generation occurs at run
time. As such, the burden of avoiding redundant states, and
ensuring that the code generators terminate is entirely on the user.
}
Run-time code generation is a generating-extension-like approach to program
specialization that produces machine code on-the-fly during program execution.
Unlike our approach to machine-code specialization, which operates on stripped
\textit{binaries} without source code or symbol-table information, run-time code
generation systems take user-annotated \textit{source code} as input and perform BTA
and generating-extension construction as part of compilation.
In the Fox \cite{leone96,PLDI:LL96} and Lancet \cite{rompf14} systems, type-level information in
the source code is exploited to produce run-time machine-code generators.
These systems avoid the state-management issues from \sectref{StateManagement} by
exploiting the availability of high-level semantic information from the source
language.
In $\rq \textrm{C}$ \cite{engler96pe}, the user implements code
generators using a DSL, and the user has the burden for avoiding
redundant states and ensuring that code generators terminate.
In contrast, \citet{klimov09} describes a run-time code
generator for Java bytecode that does not rely on information from source code.
However, \citeauthor{klimov09} can only determine
state equality for programs that do not use the heap; the approach
identifies semantically identical states based on
structural properties of Java Virtual Machine heap configurations.
JIT compilation \cite{Aycock:jit} is an example of run-time code generation in widespread use.
However, because it is performed at run-time, the emphasis is on recouping
the cost of translation, which limits the kinds of optimization techniques
that can be performed.

Turning to interpretation-based approaches, WiPER \cite{SR:Wiper} and TRIMMER
\cite{sharif18} are partial evaluators for x86 binaries and LLVM IR, respectively.
WiPER uses CodeSurfer/x86's semantic models of the 32-bit x86
instruction set to evaluate instructions.
WiPER represents states using an applicative-map-based data structure
that does not use hash-consing.
Thus, state equality is determined by directly comparing the contents
of the data structure.
\Omit{
  TRIMMER implements a non-traditional approach to partial evaluation.
  While conventional partial evaluators implicitly perform optimizations---such
  as constant folding and loop-unrolling---as a by-product of the
  partial-evaluation algorithm, TRIMMER reverses this relationship, and uses a
  pipeline of optimizations to perform partial evaluation. In particular,
  TRIMMER implements a combined constant-propagation and loop-unrolling pass. In
  effect, TRIMMER implements a more aggressive, but still sound extension of the
  aforementioned optimizations, rather than full partial evaluation, and thus
  avoids the need for a general-purpose state-management strategy. }
TRIMMER implements a non-traditional approach to partial evaluation. In
particular, TRIMMER implements an aggressive extension of LLVM's loop-unrolling
and constant-propagation passes, rather than full partial evaluation. By doing
so, TRIMMER avoids the need for a general-purpose state-management strategy.

Although our system is quite different from Fox \cite{leone96,PLDI:LL96} in most
respects, their use of \textit{pseudo-instruction macros} inspired our approach
to constructing machine-code generating extensions.
We use similar macros to produce residual assembly code, and 
extended the approach to include various other state-management actions.

\subsubsubsection{Incremental State Hashing.}
To the best of our knowledge, our work is the first application of
incremental state hashing to program specialization.
Our \texttt{fork}-based method for managing partial states was
inspired by the state-management mechanism in the EXE symbolic-execution system
\cite{exe06}. 

Model checking is a method to check properties of programs statically
by exploring the state space of a transition system.
To achieve acceptable performance, model-checking algorithms must
avoid exploring redundant states, and Rabin's fingerprinting technique
has been used to implement incremental hashing of program states
in model checkers \cite{nguyen08,mehler06}.
One of these model checkers, StEAM \cite{SPIN:LME04,mehler06}, harnesses
a VM that \emph{interprets} assembly language.
In contrast, our implementation of incremental state hashing exploits
OS-level information to apply the technique to code that \emph{executes natively},
rather than in an interpreter.

\subsubsubsection{Symbolic and Concolic Execution}
Partial evaluation bears some resemblance to symbolic execution.
In both cases,
the state space of the program is partitioned: into supplied and delayed variables in
partial evaluation, into symbolic and concrete variables in symbolic execution.
Moreover,
we use the OS-based state-management techniques previously implemented in
systems such as EXE.

However, symbolic execution differs significantly from partial evaluation in
terms of how the partitioned state space is explored.
In both symbolic execution and partial evaluation, part of the state space is
kept concrete.
In the case of symbolic execution, the non-concrete part of the state space is
represented as sets of symbolic values, which are logical formulas in some theory.
Partial evaluators, on the other hand, do not track any information about the
non-concrete part of the state space; the congruence property of the BTA
algorithm ensures that the delayed state is never needed to update supplied
values.

Symbolic execution thus attempts to construct a symbolic approximation of the
values in the symbolic portion of the state at every program point.
That is, it attempts to explore, as exhaustively as possible, precisely the
subset of the state space that a program specializer ignores.
The state-management techniques in symbolic execution are geared towards
managing the \textit{symbolic state}.
In particular, we are not aware of symbolic-execution engines that enforce any
sort of termination or state-equality properties with respect to the
\textit{concrete} portion of the state.
To perform partial evaluation with a symbolic-execution engine, one would need
to be able to determine when, for every program point, all concrete states
reachable from the starting concrete state had been reached.
There is no straightforward way to achieve this with a stock symbolic-execution tool.

\Omit{The algebraic properties
  of Rabin's fingerprinting scheme have most commonly been used to implement
  rolling-hash algorithms \cite{rabin81,karp87}. However, in \cite{rabin81}, the
  in-place update technique is described.}

\Omit{Similar approaches that exploit the
properties of modular arithmetic over fields can also be used to
construct universal hash functions, a related approach that has also
been used for in-place updates in model-checking \cite{mehler06}.}
\Omit{
Another possible approach to state hashing is to construct an $n$-ary
Merkle tree \cite{merkle89} whose leaves are hashes of memory pages,
and whose root summarizes the entire program state. However, a change
to one of the pages would thus require an update to every node on the
path from the leaf node to the root. To perform this update
efficiently, it would not suffice to retain the root only; to update a node
$N$, we would need the hash of every unchanged child, as well as the
updated child. Thus, it is necessary to store a Merkle tree for every
state currently in the worklist. Even if subtrees are shared in an
applicative fashion to reduce space usage, this approach would not eliminate
the need for leaf-to-root updates when producing an updated tree.

In contrast, to update a Rabin fingerprint, only five pieces of information
are required: the pre-state hash (a 128-bit integer);
the address of, and contents of, the pre-state changed memory;
and the address of, and contents of, the post-state changed memory.
}



\section{Conclusions and Future Work}
\label{Se:Conclusion}

This paper describes how
(i) the desire to perform specialization of machine code, using
generating extensions running natively, motivated
(ii) the development of new techniques for state management in a
program specializer.
The main challenge was that machine-code programs perform arbitrary
reads and writes to an undifferentiated address space, and for this reason,
our solution---in part---makes use of existing OS-level functionality.
Our technique is used in the generating extensions created by
\toolname\Omit{, the tool we created for specializing
stripped x86 binaries}.

It has not escaped our attention that our technique can also be used
for \emph{source-code} specialization.
In fact, we have already adapted the implementation to create
\toolnameC, a prototype generating-extension generator for C programs.

The state-management technique presented in this paper is not the only
option for a source-code specializer:
because sufficient information about a program's variables is available,
an \emph{interpreted} approach to source-code generating extensions could track
the state of memory at the level of individual variables.
However, for source-code specialization, our technique offers three advantages:
\begin{itemize}
  \item
    A generating extension for a source-code program is compiled to
    machine code, and hence specialization is performed by
    compiled---rather than interpreted---code.
  \item
    By intercepting CoW faults and incorporating changed pages into
    the hash-value for a memory state, a program specializer can easily
    support programs that use linked data structures, with no need
    to perform a mark-and-sweep traversal to capture program state.
    Thus, our technique provides a method that can be used for languages
    that are not memory-safe,
    such as C (although a conservative
    mark-and-sweep algorithm \cite{SPE:BW88} could be employed).
  \item
    The state-management implementation can be shared among program
    specializers for different languages.
\end{itemize}
As future work, we plan to develop \toolnameC further, and to investigate its
applications.



\begin{acks}                            
Supported, in part,
by a gift from Rajiv and Ritu Batra;
by \grantsponsor{GS100000003}{ONR}{https://www.onr.navy.mil/} under grants
  ~\grantnum{GS100000003}{N00014-17-1-2889} and 
  ~\grantnum{GS100000003}{N00014-19-1-2318};
and by the UW-Madison OVCRGE with funding from WARF.
The U.S.\ Government is authorized to reproduce and distribute
reprints for Governmental purposes notwithstanding any copyright
notation thereon.
Opinions, findings, conclusions, or recommendations
expressed in this publication are those of the authors,
and do not necessarily reflect the views of the sponsoring
agencies.
T.\ Reps has an ownership interest in GrammaTech, Inc.,
which has licensed elements of the technology reported
in this publication.
\end{acks}

\bibliography{df,binary,new}


\end{document}